\documentclass[a4paper]{aa}
\usepackage{amssymb}
\usepackage{graphicx}
\usepackage{natbib}
\usepackage{babel}
\usepackage{hyperref}
\usepackage{color}
\usepackage{longtable}
\def\Teff{$T_{\rm eff}$~}
\def\logg{$\log~g$~}
\def\vsini{$V$ sin $i$~}
\def\Vm{$V_{\rm m}$~}
\def\liA{$^6Li$~}
\def\liB{$^7Li$~}
\def\Li{$^7Li/^6Li$~}
\def\LTE{log N$_{\rm LTE}$(Li)~}
\def\NLTE{log N$_{\rm NLTE}$(Li)~}
\def\kmps{kms$^{\rm -1}$~}
\def\dNLTE{$\Delta_{\rm NLTE}$}
\def\vmac{$V_{mac}$~}

\begin{document}

\title{A detailed study of Lithium in 107 CHEPS dwarf stars}

\author{Ya.V. Pavlenko$^{1,2}$,
J.S. Jenkins$^{3,4}$,
O.M. Ivanyuk$^{1}$,
H.R.A. Jones$^{2}$,
B.M. Kaminsky$^{1}$,
Yu.P. Lyubchik$^1$,
L.A. Yakovina$^1$
}

\institute{Main Astronomical Observatory, National Academy of Sciences of Ukraine, Akademika Zabolotnoho, 27, Kyiv, 03143, Ukraine
\and Centre for Astrophysics Research, University of Hertfordshire, College Lane, Hatfield, Hertfordshire AL10 9AB, UK
\and Departamento de Astronom\'ia, Universidad de Chile, Casilla 36-D, Santiago, Chile 
\and Centro de Astrof\'isica y Tecnolog\'ias Afines (CATA), Casilla 36-D, Santiago, Chile\\
}

\offprints{Ya. V. Pavlenko}
\mail{email2yp\@gmail.com}

\date{}

\authorrunning{Pavlenko et al.}
\titlerunning{Lithium in 107 dwarf stars}

\abstract{
We report results from lithium abundance determinations using high resolution spectral analysis of the 107 metal-rich stars from 
the Calan-Hertfordshire Extrasolar Planet Search programme.
}{
We set out to understand the lithium distribution of the population of stars taken from this survey.
}{
The lithium abundance with account NLTE effects was determined from the fits to the Li I 6708 \AA~resonance doublet profiles 
in the observed spectra.
}{
We  find that a) fast rotators tend to have higher lithium abundances, 
b) $\log$ N(Li) is higher in more massive/hot stars, c) $\log$ N(Li) is higher in less evolved stars, i.e. stars of  lower \logg, d) stars 
with the metallicities $>$0.25~dex do not show the presence of lithium lines in their spectra, e) most of our planet hosts rotate slower, 
f) our estimate of a lower limit of lithium isotopic ratio is \Li $>$10 in the atmospheres of two SWP and two non-SWP stars.
}{
 Measurable lithium abundances were found in the atmospheres of 45 stars located at distances of 20-170 pc from the Sun, 
 for the other 62 stars the upper limits of log N(Li) 
 were computed. We found well defined dependences of lithium abundances on \Teff, \vsini, and less pronounced for the \logg. 
 In case of \vsini we see two sequences of stars: with measurable lithium and with the upper limit of log N(Li). 
 About 10\% of our targets are known to host planets. Only two SWP  have notable lithium abundances, so
we found a lower proportion of stars with detectable Li among known planet hosts than among stars without planets.
However, given the small sample size of our planet-host sample, our analysis does not show any statistically 
significant differences in the lithium abundance between SWP and stars without known planets.
}


\maketitle

\section{Introduction}
\label{_intro}

In this work we pay special attention to the determination of the lithium abundance of a sample of metal-rich dwarf stars, mainly of spectral class G. Understanding the chemical make-up of G-stars is fundamental to our understanding of star formation and stellar evolution. In many ways, G-dwarfs are key objects to enhance our understanding of our own Galactic evolution. For instance, observations indicate there are too few metal deficient G dwarfs ('G-dwarfs problem') with respect to that which could be expected from simple models of chemical evolution in our Galaxy (e.g. \citealp{sear72}, \citealp{hayw01}), as well as other bulge dominated or disk dominated galaxies (\citealp{wort96}, see more details in \citealp{caim11}).

Lithium provides an interesting special case, because it is the last of the few elements that is thought to have been produced by the Big Bang (BB). Any model of the BB can be questioned until the lithium conundrum is understood. We know from BB theory that some \liB was created shortly after the formation of the Universe, nearly 14 billion years ago \citep{oliv00}. The amount of \liB produced by the BB was comparatively small, however, and does not account for all the lithium we can see today.  A larger portion of the observed \liB is made during the Asymptotic Giant Branch phase of a small stars lifetime, i.e. C-giants and supergiants, see \cite{wall69}, \cite{abia97}, \cite{utte07}, and references therein. On the other hand, lithium is a very fragile element. Stars, which by definition must achieve high temperatures (2.5$\times$10$^6$ K)  at the base of the convective envelope, necessary for fusing hydrogen, rapidly deplete their lithium, see \cite{anth09}.

In general,  understanding the lithium abundance in the atmospheres of stars of different spectral types is a classical problem of modern astrophysics, related very acutely with the verification of many theories of the evolution of the stars, the Galaxy, and indeed the Universe.

\subsection{$^7$Li/$^6$Li}

Results based on the measurement of \Li ratios in stellar atmospheres are rather controversial. On one hand, the direct measurements of \Li in meteorites provide the 'Solar System value', where it presumably reflects the abundance that was present in the gas cloud out of which the Sun was formed \citep{ande89}. Lithium abundance in the solar photosphere is lower by a factor of 140 to that in meteorites \citep{audo83} due to the burning of lithium in nuclear reactions at the base  of the convection zone \citep{bala98}, \liA is destroyed at a lower temperature than \liB, see fig. 2 in \cite{nels93} and references therein, so the ratio of \Li in the Solar photosphere should be $\sim$ 106 \citep{borg70}. However, the meteoritic abundance ratio is found to be \Li=12.14 \citep{borg76}. \cite{chau99} reported \Li=31 $\pm$ 4 measured in the solar wind that has been implanted in a lunar soil. The ratio this low suggests that some \liA was produced by spallation reactions of high energy protons originating in solar flares with $^{16}$O and $^{12}$C atoms. Furthermore, \cite{cayr07} and \cite{ghez09} showed that the spectroscopic determination of \Li may be affected by convective motions in stellar atmosphere, adding more uncertainty when drawing conclusions from the study of the \Li ratio. On the other hand, \cite{chau99} provide some arguments suggesting that the \Li ratio has not changed significantly during the last 4.5 billion years and that a ratio $\sim$ 12 represents most gas in the solar neighbourhood.

\subsection{Lithium in exoplanet hosting stars}

Recently, many works studying the lithium abundance distribution of exoplanet hosting stars (SWP, i.e. stars with planets), find a depletion of lithium when compared to stars without planets. Generally speaking, we have a rather controversial picture here.

\cite{king97} suggested a connection between Li depletion and planet hosting after finding a difference in the abundance for the stars of the binary system 16 Cyg, one of which hosts a Jupiter-mass planet. However, the general picture is still rather controversial. On one hand, the increased rate of planet discovery in recent years has provided more material to analyse. \cite{isra04} show that SWP that have effective temperatures in the range 5600--5850 K exhibit a possible excess of Li depletion, whereas there is no significant differences in the temperature range 5850--6350 K. These results were then confirmed by \cite{take05} for a sample of 160 F, G, K dwarfs and subgiants of the Galactic disk. \citet{gonz08} and \citet{gonz10} found that the Li abundances in the atmospheres of F and G SWP with \Teff$<$5800 K are lower than those for stars without detected planets. SWP also have smaller \vsini values, which confirms the link between planet formation and stellar rotation. \citet{sous10} and \citet{delg14} showed that solar-like SWP within the temperature range 5700$<$\Teff$<$5850 K have lithium abundances that are significantly lower than those observed for field stars without any detected planets. Recently, \cite{mish16} claimed that the lithium abundances in the planet-hosting solar-analogue stars of their sample of 200 G-K dwarfs in the solar neighbourhood located at distances less than 20 pc are lower than those in stars without planetary systems. Finally, \cite{figu14} showed a significant Li depletion for SWP if a linear relation between the fundamental stellar parameters \Teff, [Fe/H], \logg, age, and Li abundance is assumed.

On the other hand, there are a number of studies that show no tendency in Li abundance decreasing in the atmospheres of SWP. \cite{luck06} found no differences between the lithium abundances of stars with and without planets. \cite{baum10} showed that for solar-like stars there are no statistical correspondence between the lithium abundances and age for stars with and without planets. Thus, the lithium abundance in stellar atmospheres is not affected by the presence of a planet near to a star. \cite{mele10} argue that the SWP with effective temperatures near \Teff$\sim$5800 K do not show anomalously low Li abundances in comparison to the stars without planets.

\cite{ghez10} analysing the behaviour of the Li abundances over a narrow range of effective temperatures (5700K$<$\Teff$<$5850 K) found subtle differences between the stars with and without planets. But the authors note that  through this temperature range various physical processes can affect the lithium depletion.

Using a compilation of 671 of their own measurements, supplemented by 1381 literature measurements of stellar parameters, \cite{rami12} show the absence of enhanced lithium depletion in SWP. The differences in the lithium abundance distribution of known SWP relative to otherwise ordinary stars appear when restricting the samples to narrow ranges of effective temperatures or mass, but they are fully explained by age and metallicity biases. In that way \cite{rami12} confirm the lack of a connection between low lithium abundance and planets.

Finally, the narrow dispersion in the metal abundances of the observed stars of the Pleiades, that have similar effective temperatures, but different lithium abundances, makes it unlikely that accretion of the hydrogen-depleted planetesimals also plays a role in the dispersion of lithium abundances \citep{wild02}.

With respect to the differences of \Li in the atmospheres of SWP and those of single stars, the situation is far from clear. On one hand, \cite{isra03} claimed the presence of a notable amount of \liA in the atmosphere of the planet host HD 82943. Yet \cite{redd02} did not find \liA in several of the eight stars that were identified in the literature as possible recipients of accreted terrestrial material, in addition to the star HD 219542 A, the planetless primary of a binary.

Thus, in spite of a number of studies devoted to this topic, there is no consensus that SWP exhibit a decreased abundance distribution when compared to stars without planets, meaning the presence of planets may, or may not, affect the abundance or processing of lithium in stellar atmospheres.

 One of the main problems of comparing planet hosts with stars without planets is to have a proper comparison sample. For example, \cite{rami12} used both, the stars without planets, and the stars that indeed may have giant planets because they do not belong to any known planet search survey. In that sense, the current paper can help on this issue since our CHEPS sample of non-hosts has also been surveyed for planets. Another advantage of this study is the homogeneity of the stellar parameters and abundances, all of them derived in the same way and with the same data. This is something that has not been considered in the past either by some works using and mixing literature values.

\subsection{Lithium formation/sink processes} 
\label{_forsink}

It is worth noting that any possible Li abundance changes of SWP, occur within the background of other large scale and time-dependent processes. Indeed, the production of lithium continues in the post-BB epochs, with a larger portion of the observed lithium being produced in the Asymptotic Giant Branch (AGB) phase of low-mass stars evolution, i.e. C-giants and supergiants. However, something is still missing from the AGB models (see \citealp{abia97}, \citealp{utte07}) and an extra mixing mechanism must be at work. The observed lithium distribution in the atmospheres of AGB stars can be explained by the circulation of material below the base of the convective envelope moving into regions where nuclear burning can happen.

We believe that canonical AGB stars produce \liB via the Cameron-Fowler mechanism (\citealp{came71}), which involves the production of beryllium deep in the hydrogen burning shell via the reaction $^4$He(3He,$\gamma$)$^7$Be and the immediate transport of it to cooler regions of the star  where $^7$Be decays to \liB, see \cite{sack95}.

The other isotope, \liA, however is made only via cosmic rays, stellar wind interactions, or in stellar flares (see \citealp{cana77}).

Recently, \cite{izzo15} revealed a 'very clear signature' of lithium speeding away from the stellar nova explosion at a speed of 2 million kilometres per hour, the first time lithium has ever been seen being produced by a nova.

Lithium can also be produced by strong stellar flares. \cite{murp90} reported convincing evidence for the existence of the (\liA)-($^7$Be) feature in a strong solar limb flare. \cite{koto96} showed that sufficient amounts of lithium might be produced in flares that could help to explain the observed lithium abundance in the Sun. \cite{livi97} reported an enhancement of the Li I 6708 \AA~ resonance feature, and particularly the presence of \liB in sunspots, and found evidence for lithium enhancement in the post-flare umbra region. It it worth noting here that the Sun is known to be a comparatively quite star (see \citealp{gurz84}). Such events may play more prominent roles in stellar atmospheres that have higher levels of magnetic activity than the Sun, or even at other stages of stellar evolution where the activity levels of stars are higher, like when they are first formed. \cite{mont98} reported the possible detection of a Li I $\lambda$ 6708 \AA~line enhancement during an unusual long-duration optical flare in the chromospherically active binary 2RE J0743+224 with a K1 III primary. During the flare, the Li I photospheric line strength gradually increased by about 40\%, and the \liA/\liB ratio, as measured by the wavelength of the Li I doublet, increased to about 10\%. Due to the complicated physics of processes in flare and post-flare stellar atmospheres, we are still far from final understanding. Determination of the local enhancement of lithium abundance in flare regions is not an easy task at all, some recent works have not found evidences of Li production in superflares, e.g. \cite{hond15}.

In general, the observed lithium abundances in stellar atmospheres at the present epoch were formed as a result of the many different processes which vary with masses of stars, initial lithium abundance at the time of formation, rotation, binarity, mass accretion/loss, activity, etc. In fact, most of the primary lithium was burning in the interiors of stars from the times of the BB. In our epoch \liA is observed in the spectra of some objects, however, from the two stable isotopes of lithium, \liB is more abundant (\citealp{choi13}).

The layout of the manuscript is as follows, in Section \ref{_obs1} we provide information about the stars of our sample and the observed spectra, in Section \ref{_pro} we discuss the details of the LTE and NLTE procedures of our  lithium abundance determination in the framework of both approaches. Sections \ref{_res} and \ref{_res1} contain the description of our measured lithium abundance results. In the Section \ref{_dis} we discuss the comparison of our results with previously published works and we summarise our findings.

\section{Observations and data acquisition} 
\label{_obs1}

We used the observed spectra obtained in the framework of the Calan-Hertfordshire Extrasolar Planet Search (CHEPS) programme \citep{jenk09}. The program was proposed to monitor samples of metal-rich dwarf and subgiant stars selected from Hipparcos, with $V$-band magnitudes in the range 7.5 to 9.5 in the southern hemisphere, in order to search for planets that could help improve the existing statistics for planets orbiting such stars, particularly the smallest possible of this cohort.

The secondary selection criteria for CHEPS was based on selecting inactive (log$R'_{\rm{HK}}\le$-4.5\,dex) and metal-rich ([Fe/H]$\ge$+0.1\,dex) stars by the analysis of high-resolution FEROS spectra \citep{jenk08,murgas13} to ensure the most radial velocity stable targets, and to make use of the known increase in the fraction of planet-host stars with increasing metallicity, mentioned above. Furthermore, SIMBAD (simbad.u-strasb.fr/simbad) does not indicate our stars are members of binary systems, however we have discovered a number of low-mass binary companions as part of the CHEPS program (e.g. \citealp{jenk09}; \citealp{pant17}).

 All stars in our work were observed with the HARPS spectrograph \cite{mayor03} at a resolving power of 115000, and since the spectra were taken as part of the CHEPS program, whose primary goal is the detection of small planets orbiting these stars, the SNR of the spectra are all over 100 at a wavelength of 6000 \AA. The 107 stars in this work are primary targets for CHEPS, however, there are additional targets that have been observed with CORALIE and MIKE that we have not included in this work to maintain the homology of our analysis, specifically these instruments operate at significantly lower resolution than HARPS.

Thus far the CHEPS project has discovered 15 planets \cite{jenk17} and a number of brown dwarf and binary companions \citep{vines17}. The high-resolution and high-S/N of the CHEPS spectra from HARPS allows \cite{ivan17} the study of chemical abundances like Na, Mg, Al, Si, Ca, Ti, Cr, Mn, Fe, Ni, Cu and Zn in the atmospheres of metal-rich dwarfs. Gravities in the atmospheres of subgiants are lower in comparison to dwarfs, however, we carried out the same procedure for the lithium abundance determination for the stars of both groups. Differential analysis of the results allows us to investigate the effects of gravity, effective temperature, etc. on the present stages of evolution of our stars.

\section{Procedure}
\label{_pro}

\subsection{Abundances analysis}

 We used the  stellar atmospheric parameters (effective temperatures \Teff, gravities \logg, microturbulent velocities \Vm, rotational velocities \vsini and abundances of Na, Mg, Al, Si, Ca, Ti, Cr, Mn, Fe, Ni, Cu, and Zn determined for the 107 stars of CHEPS sample by \citet{ivan17}, who used our procedure of finding the best fit of the synthetic absorption line profiles to the observed spectra using ABEL8 program \citep{pavl17}. We used the model atmospheres computed by \cite{ivan17} using SAM12 program \citep{pavl03}. Model atmospheres, synthetic spectra were computed for the same set of input parameters. This became the basis for our lithium calculations that we detail below.

 Our LTE analysis of lithium abundances is based on the fits of our synthetic spectra to the observed Li line profiles. 
For the lithium abundance determination procedure we assume that both the instrumental broadening and macroturbulent broadening have Gaussian profiles.
We adopt that a macroturbulent velocity \vmac distribution in the atmospheres of our stars that is similar to the case of the solar atmosphere. 
For the case of the solar spectrum, the macroturbulent velocities
mean that the measured widths of solar lines corresponds to a resolution,
$R$ = 70K at 6700 \AA. This
formal resolution is limited by the presence in the solar atmosphere of macroturbulent
motions with \vmac = 1 - 2.6 km/s, see \cite{pavl12} and references therein. 
We use this value of $R$ for all our spectra. Generally speaking, \vmac varies with depth in the atmosphere, depending on
the physical state of the outer part of the convective envelope of a star.
We adopt a solar-like model of macroturbulence for all stars in our sample.
It is worth noting that in the case of slow rotators, the determination of \vsini and \vmac is a  degenerate problem. On the other hand,
instrumental broadening, rotational broadening and broadening by macroturbulent velocities do not affect the 
integrated intensity of absorption lines.
Nevertheless, fixing the macroturbulent velocity simplifies the procedure. 

Furthermore, for the case of our \Li analysis we were particularly interested in gaining the most accurate computations 
of the Li line profiles possible. Therefore in the \Li analysis we used more complicated line profiles for 
the macroturbulent velocity distribution, see Section \ref{_res1}. 

To obtain more accurate fits to the observed spectrum we used the \vsini parameter
to adjust our fits to the observed line profiles.
Rotational velocity was varied around the values found by
\cite{ivan17} to obtain the best 
fit to the observed Li lines. In most cases, differences of our \vsini with \cite{ivan17} 
does not exceed $\pm$0.5 \kmps.

Our simplified model does not consider 
changes of \Vm and \vmac with depth, differential rotation 
of stars, presence of spots, the effects of magnetic activity, etc.
We believe that taking account of these effects should not significantly alter our results, at least as relates to the 
lithium abundance determination. The detailed modelling
of any of the listed effects we outline here requires very sophisticated analysis that is beyond the framework 
of our paper.

The effects of line blending were treated explicitly, whereby only well fitted parts of the Li line profile were used in our analysis. This approach allowed us to minimise effects of blending by lines of other atoms and even CN, which depends on the unknown yet C and N abundances. In this work abundances of C and N and other non-analysed elements by \cite{ivan17} were scaled following [Fe/H]. The updated atomic line list VALD-2 (\citealp{kupk99}) was used in our synthetic spectra computations, see \cite{yako11}. \footnote{Our list of atomic lines for the 6682-6742 \AA~spectral region is available on ftp://ftp.mao.kiev.ua/pub/yp/2017/Li6708/val67-08.5aug2.}

\subsection{LTE analysis}

Firstly, we determined the lithium abundances in the atmospheres of the stars in our sample in the framework of the thermodynamic equilibrium (LTE) approach following the \citet{pavl12,pavl17} algorithm. LTE model atmospheres and synthetic spectra were computed with the SAM12 and WITA6 programs, respectively, (see \citealp{pavl97, pavl03}). The fine details of our procedure for the determination of abundances assuming LTE are described in \cite{ivan17}.

\begin{figure}
\centering
\includegraphics[width=55mm]{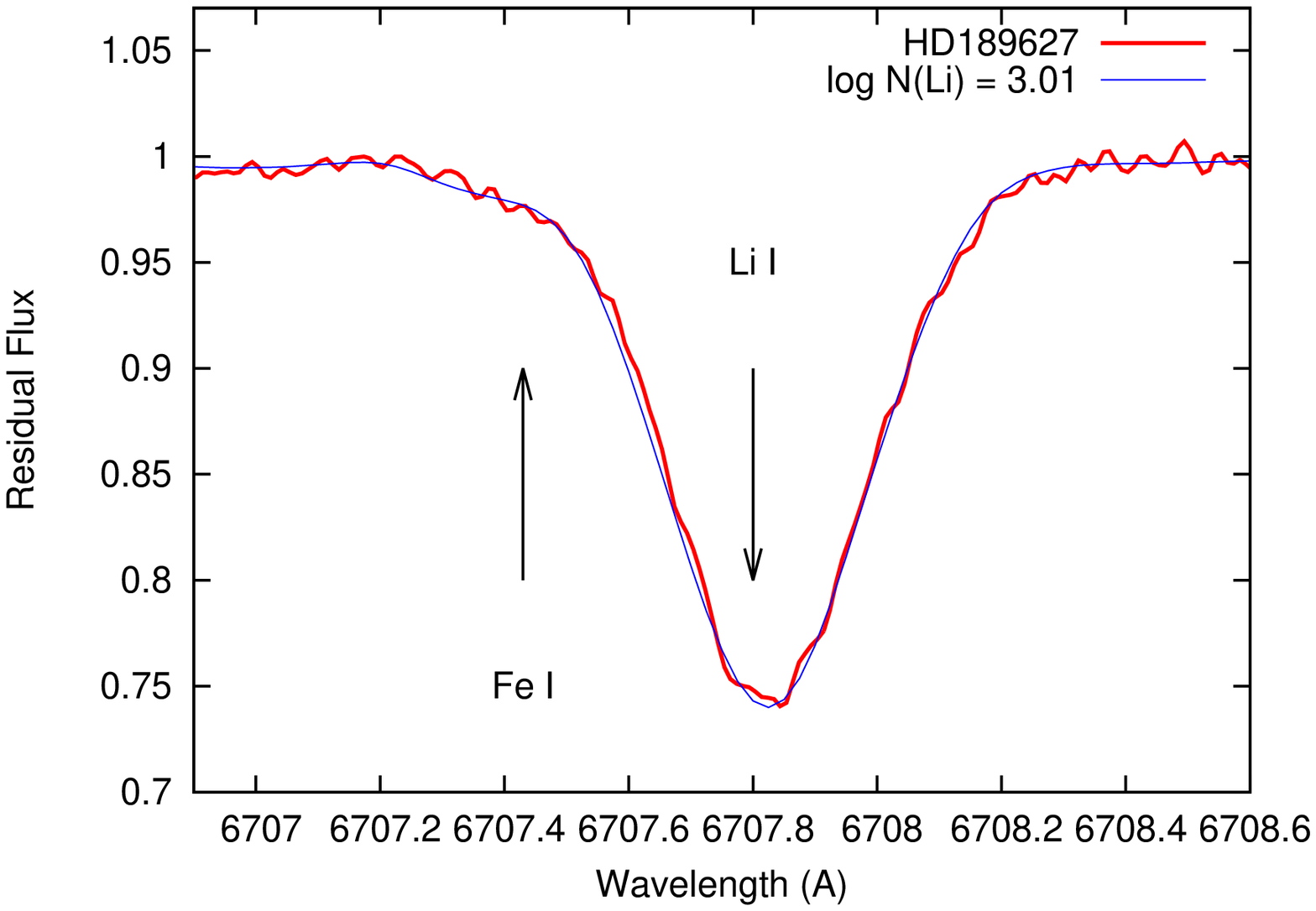}
\includegraphics[width=55mm]{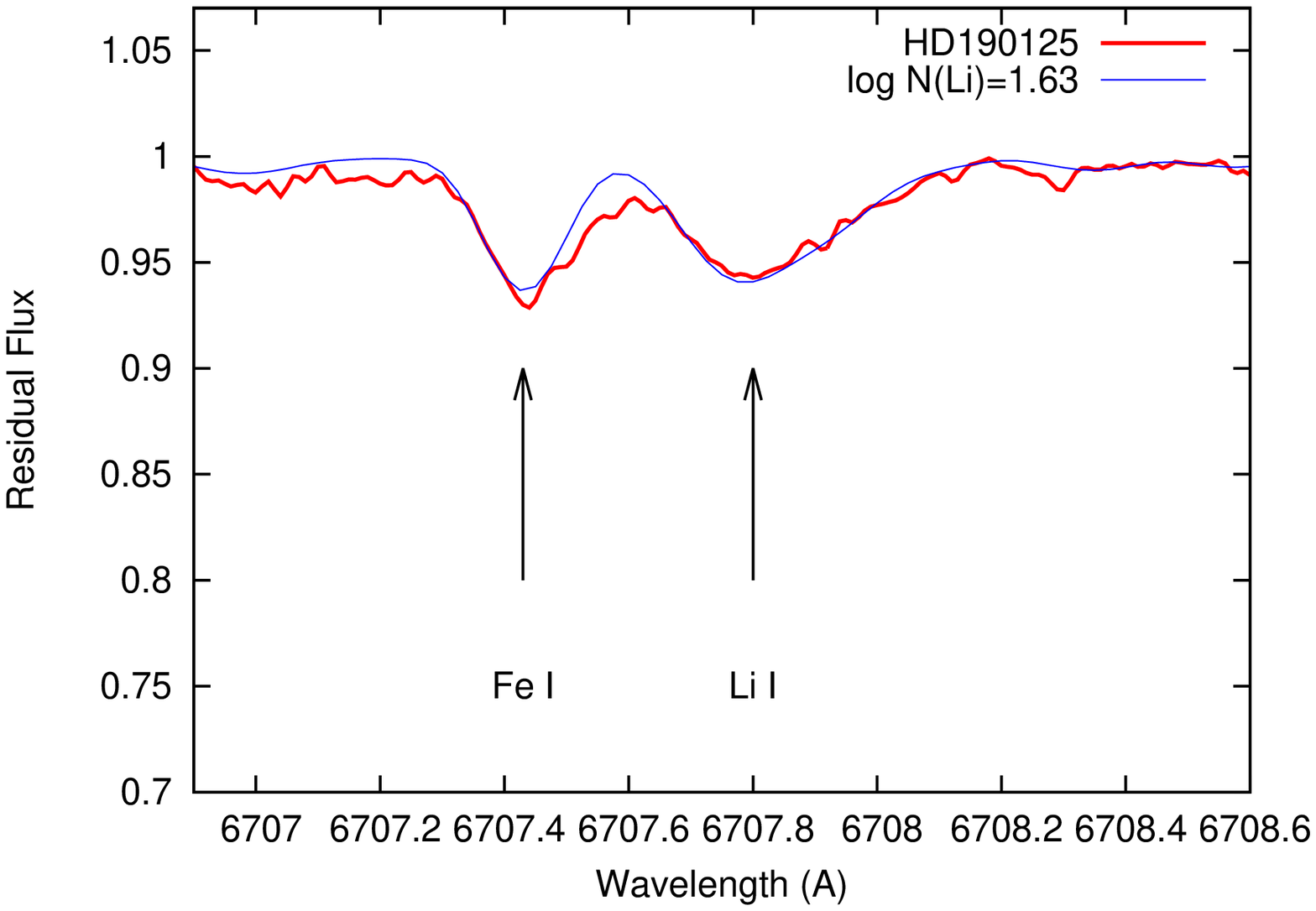}
\caption{Fits of the synthetic spectra to the observed lithium line at 6708 \AA~for the spectrum of HD 189627 ({\it top}) and HD 190125 ({\it bottom}). The contribution of the Fe I line at 6707.473 \AA~ in the formation of the blend is clearly seen in both cases, and is particularly noticeable in the spectra of HD 190125. The used input parameters for these computations are given in the Table \ref{_lires10}.
\label{_liex}}
\end{figure}

Our LTE analysis of the Li abundances is mostly based on the fits of our synthetic spectra, computed for log N(Li) from 0.0 to 3.5 with an abundance step of 0.05 dex, to the observed profiles of absorption lines of the resonance lithium doublet at 6708 \AA. We used the fits to observed profiles of the Li I line in this way to minimize the effects of blending by other lines, which is of particular importance here due to the enhanced metallicities of our target sample. In Fig. \ref{_liex} we show fits of our synthetic spectra to the observed lithium lines in the spectra of HD 189627 and HD 190125. These two plots represent cases of the highest and one of the lowest lithium abundances measured in the atmospheres of our sample. The Li resonance doublet is blended with the Fe I line at 6707.43 \AA, and therefore to get the most accurate Li abundance we used the not blended part of the broad Li resonance doublet profile.

\begin{figure}
\centering
\includegraphics[width=55mm]{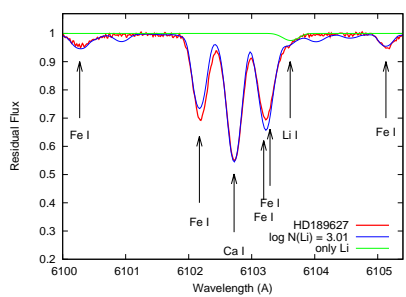}
\caption{Fits to the spectral region across the Li subordinate 6103 \AA~line for the star HD 189627. The red curve is the observed spectra, the blue curve is 
the best fit model, and the green curve shows the profile of the Li I line computed for log N(Li)=3.00.  The used input parameters for these computations are 
given in the Table \ref{_lires10}. 
\label{_li6103}}
\end{figure}

In the case of late type stars the subordinate lithium lines at 6103 \AA~and 8126 \AA~ can be used to measure the lithium abundance, if they are strong enough in the extracted high resolution spectra, which is not always the case.  Unfortunately, the Li I subordinate triplet at 6103 \AA~ is too weak even in the spectrum of the most Li rich stars from our sample, see Fig. \ref{_li6103}. Blending iron lines are too strong here, therefore, we do not use 6103 \AA~ Li I line in our analysis.

Another subordinate doublet is located beyond the spectral range covered by HARPS. In comparison to the work of \cite{ivan17}, we carried out the lithium abundance measurements for the fixed model atmosphere structures. Indeed, direct experiments showed that changes of the lithium abundance that are within a reasonable range i.e. log N(Li)$<$3.5, cannot affect our model atmosphere structures across all ranges of our sample's effective temperatures and surface gravities.

\subsection{NLTE abundances}

To carry out the NLTE analysis for a 20-level Li atom model, we followed the procedure described in \cite{pavl94} and \cite{pavl96}. We used the same opacity source list and ionisation-dissociation approach as for the model atmospheres and synthetic spectra procedures, see \cite{ivan17}.

Here it is pertinent to note a few aspects of our NLTE calculation:

Our 20-level lithium atom model allows us to consider the interlocking of lithium lines (transitions) more appropriately. The ionisation equilibrium of lithium is formed by the whole system of the bound-free transitions, where transitions from/to the second level play the main role. The radiation field in bound-free transition computations is very important in the modelling of lithium lines without LTE. The effectiveness of the overionisation of lithium depends directly on the mean intensities of the radiation field in the blue part of the spectrum (\citealp{pavl89,pavl91}). Lithium lines are treated here explicitly, i.e. as multiplets. Basically, radiation transfer in the frequencies of several multiplet lines should differ from the case of one single (strongest) line. We account the contribution of molecular line absorption (CH, CN and other) in the frequencies of bound-bound and bound-free transitions of Li I atom. In principle, this should reduce the probability of photon losses from the atmosphere, i.e. it directly affects the processes of radiative transfer.

\subsection{NLTE curves of growth, synthetic spectra and abundances}

The NLTE computations were performed by using a modified version of the program, which is described in detail in \citep{pavl99}. The following is an account of the NLTE abundance corrections we have performed in this work:

$\--$ Firstly, we compute LTE curves of growth, accounting for the multiplet structures of the lithium 6708 lines.

$\--$ A self-consistent system of statistical balance, together with the radiation field transfer equation (NLTE problem) was solved using the modified LINEAR2 program \citep{auer76}.

$\--$ Using the information acquired on the NLTE populations of the lithium levels, we compute NLTE curves of growth.

$\--$ Finally, the shift between LTE and NLTE curves of growths provide the NLTE abundance correction for Li.

\section{Results}
\label{_res}

\begin{figure*}
\centering
\includegraphics[width=0.48\linewidth]{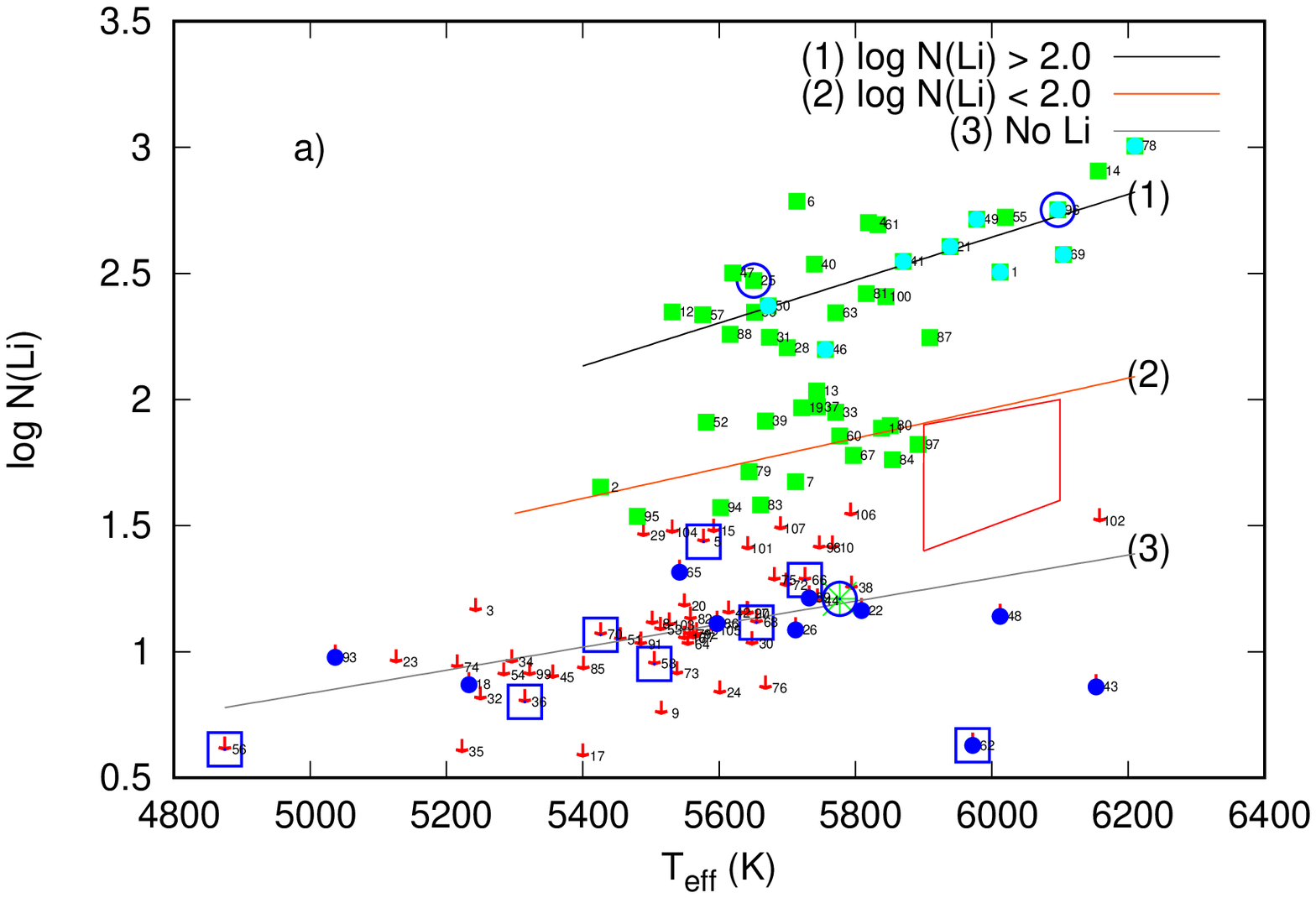}
\includegraphics[width=0.48\linewidth]{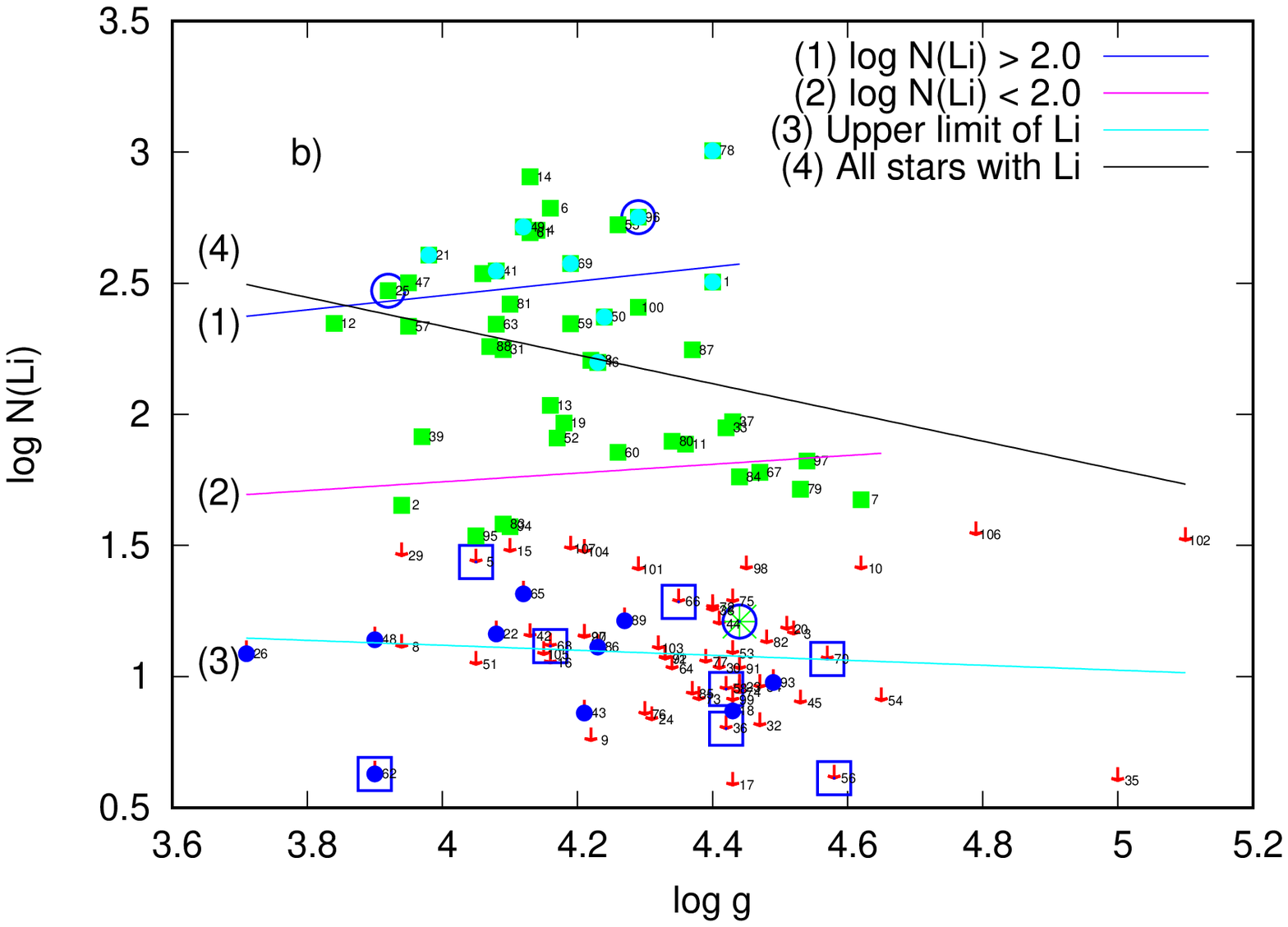}
\includegraphics[width=0.48\linewidth]{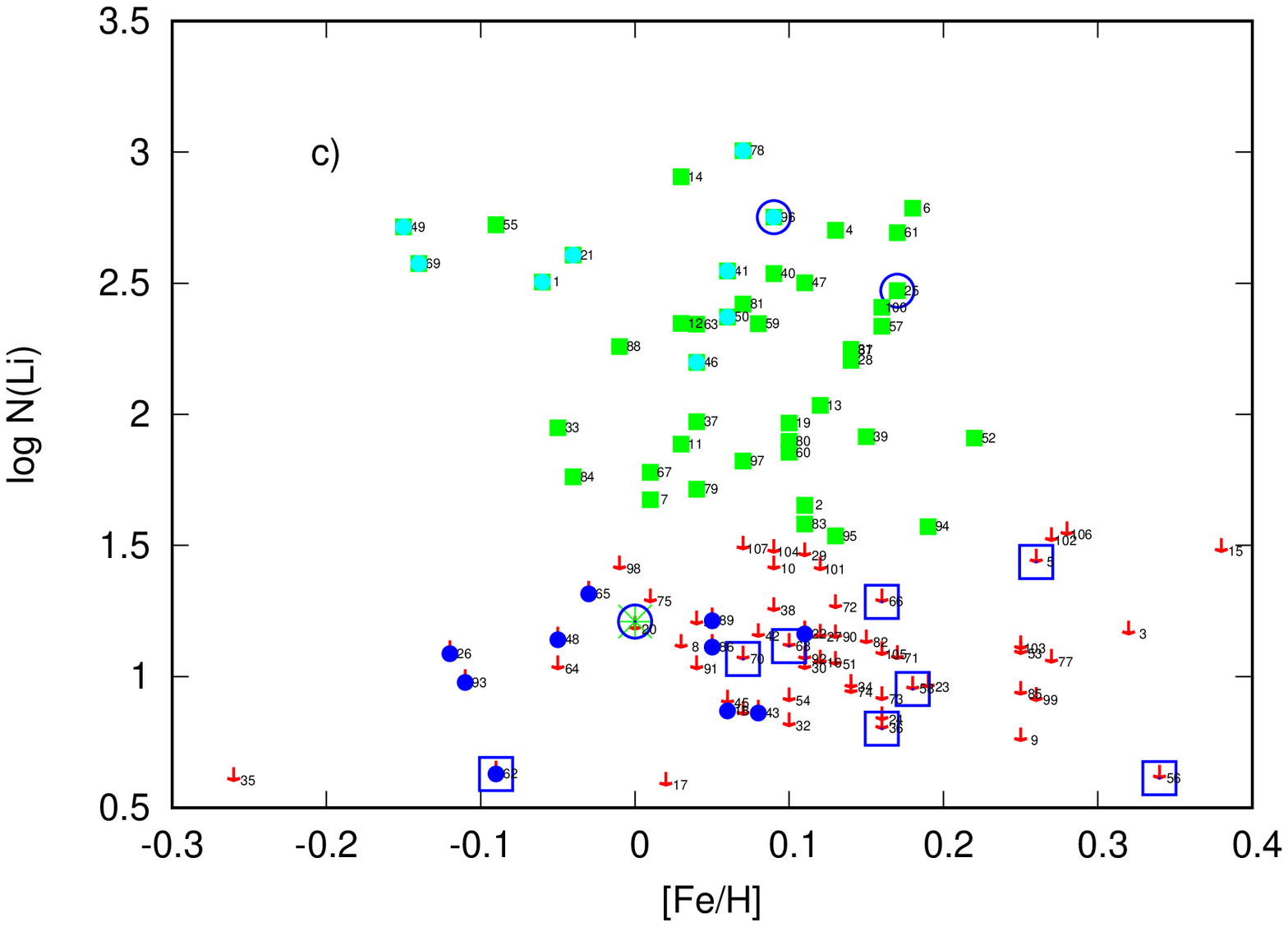}
\includegraphics[width=0.48\linewidth]{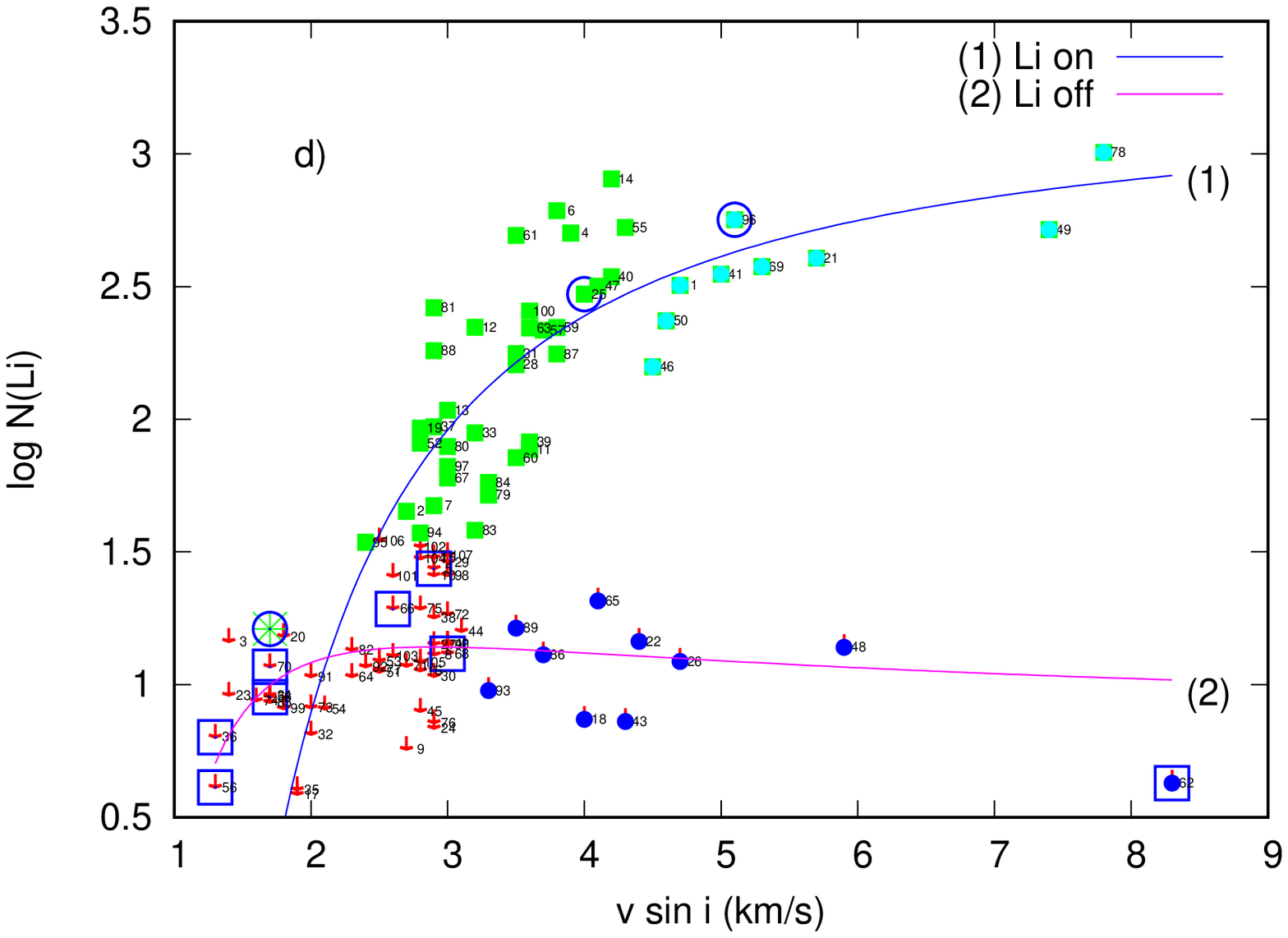}
\includegraphics[width=0.48\linewidth]{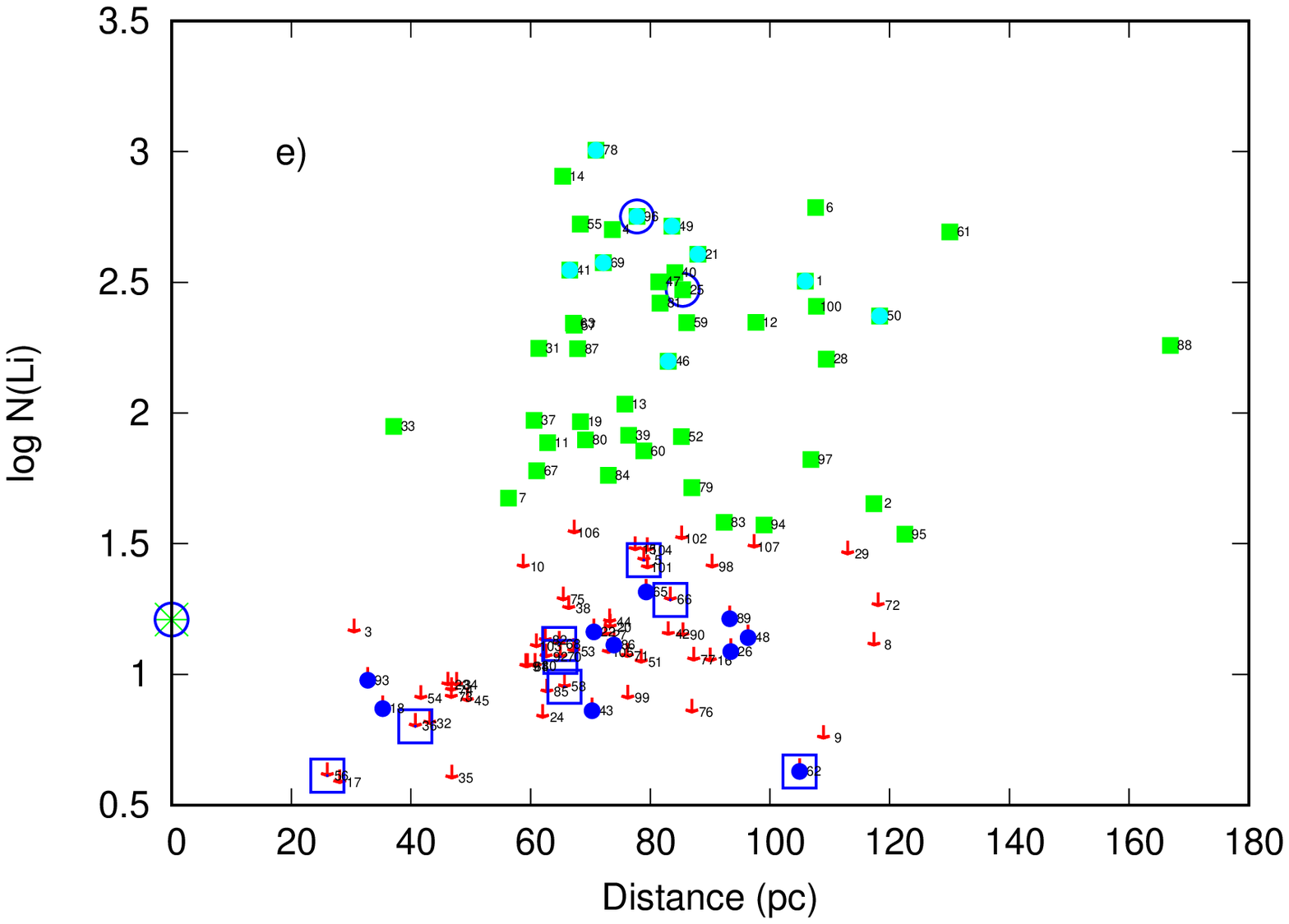}
\includegraphics[width=0.48\linewidth]{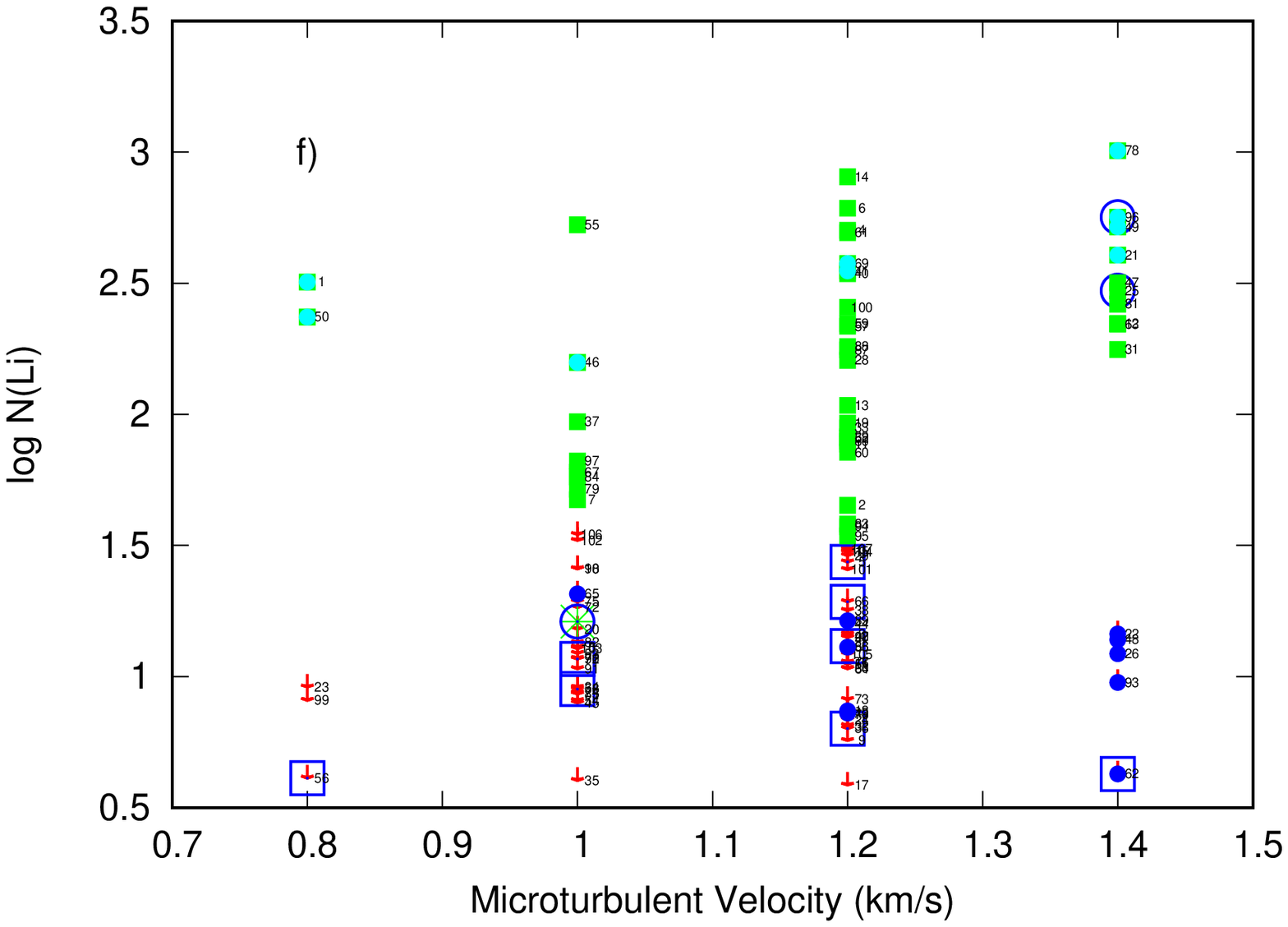}
\caption{Abundances of lithium vs. \Teff (a), \logg (b), [Fe/H] (c), \vsini (d), distance (e), and \Vm (f), all shown by filled boxes. Cyan squares and blue circles mark fast rotators with high lithium and without notable lithium, respectively. Upper limits for the lithium abundances in the spectra of stars of our sample are marked by arrows. Dwarf stars hosting substellar components (see \citealp{jenk17}) with and without strong lithium lines in the spectra are marked by large open circles and squares, respectively. The polygon in (a) bounds the approximate extent of the lithium desert and it is drawn here only to guide the eye, as in \cite{rami12}. The NLTE lithium abundance in the Sun \citep{pavl89} is shown by the green asterisk in the blue ring. The numeration of stars follows the ordering os stars in the Table \ref{_age}.
\label{_vs}}
\end{figure*}

 We computed Li I lines at 6708 \AA~as a blend created by multiplet transitions. LTE and NLTE abundances, as well as NLTE lithium abundances corrections $\Delta_{\rm NLTE}=$log N$_{\rm NLTE}$(Li)-log N$_{\rm LTE}$(Li), computed for all stars with noticeable Li line at 6708 \AA, are shown in Tables \ref{_lires11}, \ref{_lires10}, \ref{_lires00}. We found the largest NLTE lithium abundance correction to be 0.113 for HD 7950, one of the coolest stars in our sample. For the other stars $\Delta_{\rm NLTE}$ are lower.

As was shown by \cite{pavl94} for the case of G-K stars with solar abundances, the significance of NLTE effects depends upon the line strength. Weak Li lines are affected by the overionization processes, therefore, their departure corrections tend to be positive ($\Delta_{\rm NLTE}>$0), whereas lines of intermediate strength form in atmospheric layers where the source function $S_\nu(\tau_{\rm NLTE}\approx 1)$ is less than the LTE Plank function $B_\nu(\tau_{\rm {\rm LTE}}\approx 1)$, and NLTE cores of the resonance doublet become stronger than the LTE ones. As a result, NLTE abundance corrections $\Delta_{\rm NLTE}$ become negative, like we see in the case HD 189627, see Table \ref{_lires10}.

\subsection{Sensitivity of the lithium abundance on the input parameters}

To provide fits to the observed profiles of the Li I resonance doublet lines, we computed synthetic spectra using the conventional input data: effective temperature \Teff, gravity \logg, mictoturbulent velocity \Vm. All these parameters were previously computed to high precision by \citep{ivan17}. Uncertainties in the input data affect the final result, i.e. lithium abundances, see Table \ref{_deli}. In this table we also show the response of the lithium abundance to changes in \Teff, \logg, \vsini for the cases of lithium lines with different intensities in the observed spectra.

Interestingly, the computed changes of lithium abundances are of the same order for strong and weak lithium lines. Lithium is an element of low ionisation potential, therefore for its abundance determination we should use the most accurate \Teff determination. Li I line abundance calculations are not sensitive to the changes of \logg and \Vm at the level of $\pm$0.2~dex, and $\pm$0.2 \kmps, respectively. These parameters were determined using pre-selected 'good' Fe I and Fe II lines by \cite{ivan17} with an accuracy of $\pm$0.2 in both parameters, so the main effect for our lithium results is determined by the \Teff choice, where the photometric colours provide an accuracy of $\pm$50 K for \Teff. Therefore, we conclude that our accuracy in the lithium abundances due to the uncertainties in \Teff, \logg, \Vm does not exceed 0.05 dex.
 
 Another source of lithium abundance errors might be the continuum level uncertainties. However, our spectra are of a good quality with S/N $\sim$100 and lithium resonance doublet is located in a comparatively low opacity spectral region, i.e. not crowded by lines of other elements. Nevertheless, after the determination of lithium abundances by fitting to the observed spectra, all fits were checked visually; this allows the refinements for a few cases of extremely strong and weak lithium lines. Furthermore, we fitted only non-blended and well defined parts of lithium resonance doublet, this reduces the impact of uncertainty in the continuum level is lower in comparison to the equivalent width method. Based on our experiments adding noise to spectrum and continuum level re-determination we believe that the uncertainty in the lithium abundance caused by mistakes in the continuum level does not exceed 0.02 dex.

Due to the low potential of ionisation of lithium (5.26 eV) its lines become stronger in the spectra of stars of lower \Teff for the same lithium abundance. As a result, our upper limits of lithium abundances decrease with lowering \Teff.

\begin{table*}
\caption{Lithium abundance vs. changes of input parameters. Numbers in brackets show the \vsini used to adjust the fit of the synthetic spectra to the resonance lithium doublet profile, (see \citealp{ivan17}).}
\label{_deli}
\begin{tabular}{c|c|cc|cc|cc|}
\hline\hline
&&&&&&& \\
log N$_{\rm LTE}$(Li) & Star & \multicolumn{2}{|c|}{\Teff(K)}               &  \multicolumn{2}{|c|}{\logg} & \multicolumn{2}{|c|}{\Vm (km/s)}  \\  
          &      &   -50    &  +50     &   -0.2     &  +0.2    &    -0.2    &  +0.2   \\    
\hline
&&&&&&& \\

 0.98 &  HD 220981 & +0.05 (5) & -0.05 (5) &    0.0 (5) &  0.0 (5) &    0.0 (5) & 0.0 (5) \\ 
 1.00 &  HIP 66990 & +0.05 (4) & -0.05 (4) &    0.0 (4) &  0.0 (4) &    0.0 (4) & 0.0 (4) \\ 
 1.02 &  HD 49866  & +0.05 (4) & -0.05 (4) &    0.0 (4) &  0.0 (4) &    0.0 (4) & 0.05 (5) \\ 
                   
 1.48 &  HD 221954 & +0.05 (4) & -0.05 (4) &    0.0 (4) & 0.0  (4) &    0.0 (4) & 0.0 (5) \\ 
 1.50 &  HD 193995 & +0.05 (5) & -0.05 (5) &    0.0 (5) & 0.0  (5) &    0.0 (4) & 0.0 (5) \\ 
 1.54 &  HD 7950   & +0.05 (4) & -0.05 (4) &    0.0 (4) & 0.0  (4) &    0.0 (4) & 0.0 (4) \\ 
                   
 1.90 &  HD 78130  & 0.00 (4) & -0.05  (5) &    0.0 (5) & 0.0  (5) &    0.0 (5) & 0.0  (5) \\ 
 1.96 &  HD 19493  & +0.05 (4) & -0.05 (3) &    0.0 (4) & 0.0  (4) &    0.0 (3) & 0.0  (4) \\ 
 2.17 &  HD 56957  & +0.05 (5) & -0.05 (5) &    0.0 (5) & 0.0  (5) &    0.0 (5) & 0.0  (5) \\  
                   
 2.43 &  HD 101348 & +0.05 (6) & -0.05 (5) &    0.0 (5) & 0.0  (5) &    0.0 (5) & 0.0   (6)  \\ 
 2.48 &  HD 6790   & +0.04 (6) & -0.01 (6) & +0.01 (6) & +0.01  (6)&    0.01 (6) & 0.01   (6)  \\ 
 2.55 &  HD 90520  & 0.0   (5) & -0.01 (6) &    0.0 (6) & 0.0  (6) &    0.0 (6) & 0.0   (6)  \\ 
                   
 2.75 &  HD 10188  & +0.05 (5) & 0.0   (5) &    0.0 (6) & 0.0  (5) &    0.0 (5) & 0.0   (5)  \\ 
 2.90 &  HD 19773  & 0     (5) & -0.05 (5) &    -0.05 (4) & 0.0 (5)&   -0.05 (4) & 0.0   (5)  \\ 
 3.01 &  HD 189627 & 0.05 (9) & -0.05  (9) &    0.0 (9) & 0.0  (9) &    0.0 (9) & 0.0   (9)  \\ 
&&&&&& \\
\hline
\end{tabular}
\end{table*}

\subsection{log N(Li) vs. \Teff}

Our results for the CHEPS sample \citep{jenk17} agree well with \cite{lope15} and \cite{rami12} (see Fig. \ref{_vs}a and fig. 4 in \cite{lope15}). Lithium abundances increase toward the high temperature edge of our sample, it reflects the decrease of effectiveness of Li depletion in stars of higher mass.  Cooler stars have deeper convective envelopes which allow for Li to get into hot enough regions for processing to occur, see pioneering works by \cite{boes86}, \cite{boes188}, \cite{deli90} and \cite{thor93}.

Interestingly, we did not find the stars inside the region of the so called lithium desert located approximately at $5900<$\Teff$<6200$ K, log N(Li)$<$2. One of the hottest stars in our sample, i.e. HD 189627, shows the highest Li abundance log N$_{\rm NLTE}$(Li)=3.01$\pm0.05$. On the opposite site of the measurable dependence log N(Li)=f(\Teff), we see the coolest star in our sample, HD 128356 with log N$_{\rm NLTE}$(Li)=0.61, a value that corresponds to the measured values for the coolest stars in the sample of \cite{rami12}.

We divided our sample of stars with notable lithium lines in their spectra into two bins, with abundances of log N(Li)$>$2 and log N(Li)$<$2. Linear approximations of log N(Li)=$a+b\times$\Teff were found for these two populations, as well as for the stars that had only upper limits determined, and are shown in Fig. \ref{_vs}. The parameters $b$=0.00045$\pm$0.00011, 0.00085$\pm$0.00018, and 0.00060$\pm$0.00024 were found for the stars with the lithium abundances log N(Li)$>$2, log N(Li)$<$2, and 'without notable Li' cases, respectively. Our slopes $b=\partial(\rm logN(Li))/\partial(T_{eff})$ are always positive for all of these bins. Here it looks like we see the results of Li depletion in the stars formed in three different epochs. If this is correct, then the Sun with its low Li abundance belongs to the oldest group of stars.  Interestingly, the SWP with high Li abundance follow the fit of Li-rich stars and the SWP with upper limits follow the fit for the stars with upper limits.

\subsection{log N(Li) vs. \logg}

In Fig. \ref{_vs}b we show the lithium abundance decreasing in the atmospheres of stars of higher \logg. The effect can be interpreted either as a result of stronger lithium depletion in the atmospheres of older stars, i.e. stars of higher \logg, or due to enhanced mixing in stars with deeper convective envelopes.  Alternatively, those stars with lower \logg values would be slightly evolved and thus their \Teff at the main sequence was higher, therefore they didn't destroy so much Li due to their thinner convective envelopes.

Interestingly, the star with the highest measured lithium abundance, i.e. HD 189627 with a log N(Li)=3, does not show any peculiarity in \logg. Likely, the reason for the high lithium in its atmosphere is not related to this parameter, see Section \ref{_vsinili}.

The two groups of stars with log N(Li)$>$2 and log N(Li)$<$2 may be localised in Fig. \ref{_vs}b, as well as in other panels of the figure. Again, we approximated the dependence of log N(Li) vs. \logg by a linear function parametrised as log N(Li)=$a+b\times$\logg, and we found values of $b$=.27$\pm$0.32 and 0.17$\pm$0.16 for the two cases. Also, for all stars with and without Li, we obtained values of $b$=-0.54$\pm$0.32 and -0.09$\pm$0.12, respectively. We note that we clearly obtained a lower degree of correlation between log N(Li) and \logg for the split sample of stars with Li.

\subsection{log N(Li) vs. [Fe/H]}
\label{_life}

First of all, we do not find a notable lithium abundance in the atmospheres of the most metal-rich stars, i.e. dwarfs with [Fe/H]$>$0.25. Lithium is depleted substantially (log N(Li)$<$0.0) in the atmospheres of at least 10 stars when compared to stars toward the lower metallicity range, where half of the stars of our sample have measurable Li abundance. The lack of Li in the range [Fe/H]$>$0.25~dex may be explained by internal and external reasons. Due to higher opacities in their interiors, the most metal-rich stars may have more developed convective envelopes. Alternatively, these stars lost their lithium at the stage of formation of their exoplanetary systems (planets yet to be detected), as was suggested by \cite{gonz15} who present a table of SWP and a deficit of lithium paired with very similar stars lacking planets, extending the recent similar results of \cite{delg14}.

 The depletion of Li at high [Fe/H] was also shown by \cite{delg14}, and as the authors explained, this can be caused by the deeper convective envelopes expected for higher opacities (although some studies don't agree with that, see \cite{pins01}. However, it is worth noting that such depletion can also be caused by the Galactic chemical evolution of Li, see \cite{delg15} and \cite{guig16}.

\subsection{log N(Li) vs. \vsini}
\label{_vsinili}

To begin, SWP of all types rotate more slowly \citep{gonz15}, normally explained due to biases in the detection of these systems since slowly rotating stars are less complicated for the analysis in general. We show the effect in Fig. \ref{_vs}d, in which a majority of SWP are comparatively slow rotators with \vsini $\leq$ 3 \kmps. However, lithium is observed mainly in SWP having notable \vsini.

 In Fig. \ref{_vs}d we highlight the high and low lithium abundance 'tails' which become visible starting at 3 \kmps at low abundances and at 4.5 \kmps at high abundances. Interestingly, one of the most lithium rich stars in our sample, HD 189627, shows the largest \vsini=7.8 \kmps, and the intrinsic $V_r$ may be even higher. On the other hand, another star of our sample HD 147873 is a fast rotator with \vsini=8.3 \kmps, but this SWP does not show any notable Li features in its spectrum.

We found an approximation of log N(Li) vs. \vsini for both our subsamples of stars that have strong lithium and those without: log N(Li)=$a+b$/\vsini$+c$/(\vsini)$^2$, see Fig. \ref{_vs}d, and Table \ref{_t2}. Here we see the strong correlation between stellar rotation and lithium abundance.

\begin{table}
\caption{Coefficients of the approximation of the dependence of log N(Li) vs. \vsini.}
\label{_t2}
\begin{tabular}{c|ccc}
\hline\hline
&&&                                                                          \\
     a case      &        $a$        &        $b$        &        $c$        \\
&&&                                                                          \\
\hline
&&&                                                                          \\
Li on            &  3.26 $\pm$ 0.54  & -2.23 $\pm$ 4.07  & -5.00 $\pm$ -7.48  \\
                 &                   &                   &                   \\
 Li upper        &                    &                   &                    \\ 
 limit           &  0.84 $\pm$ 0.23    &  1.68 $\pm$ 1.00  & -2.44 $\pm$ 1.07  \\\hline
\end{tabular}
\end{table}

\subsection{log N(Li) vs. distance}

The majority of our stars with noticeable lithium in their atmospheres are located in the volume of space between 20-120 pc around the Sun. The lack of stars at distances larger than 120 pc can be explained due to the CHEPS selection mentioned above, since these stars are relatively faint for the magnitude range of the survey. Most notably here, only one star from the sample located within 60 pc of the Sun shows a log N(Li)=2, meaning it looks like the Sun lies in a lithium depleted part of Galaxy, if our metal-rich sample is representative of the sample of stars in the solar vicinity. Our stars are located at larger distances than \cite{mish16}, and most do not show notable lithium at all. 
 Our results show that more representative, in the sense of distances from the Sun, larger sample should be used for more robust conclusions.

\begin{figure*}
\centering
\includegraphics[width=0.48\linewidth]{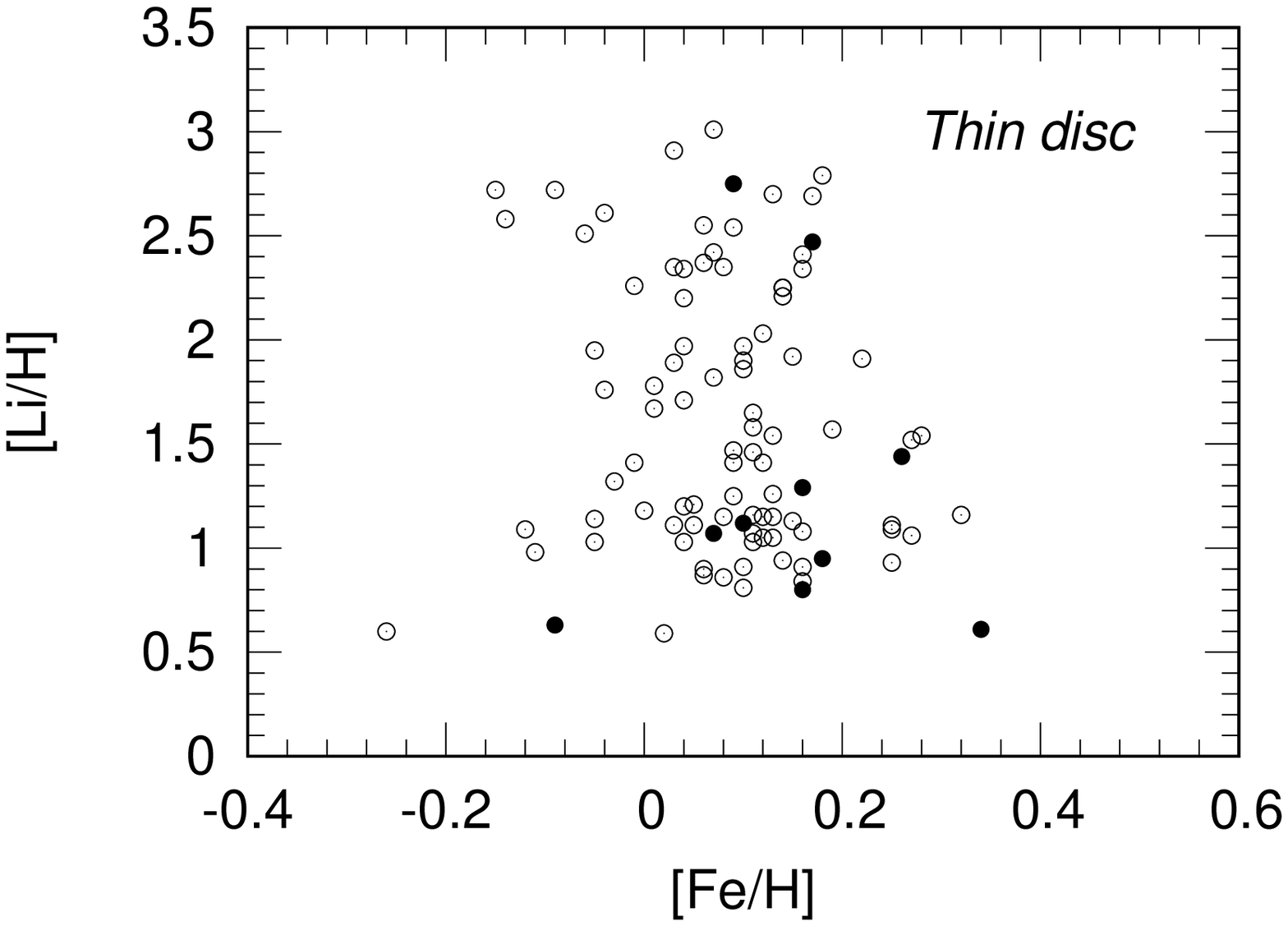}
\includegraphics[width=0.48\linewidth]{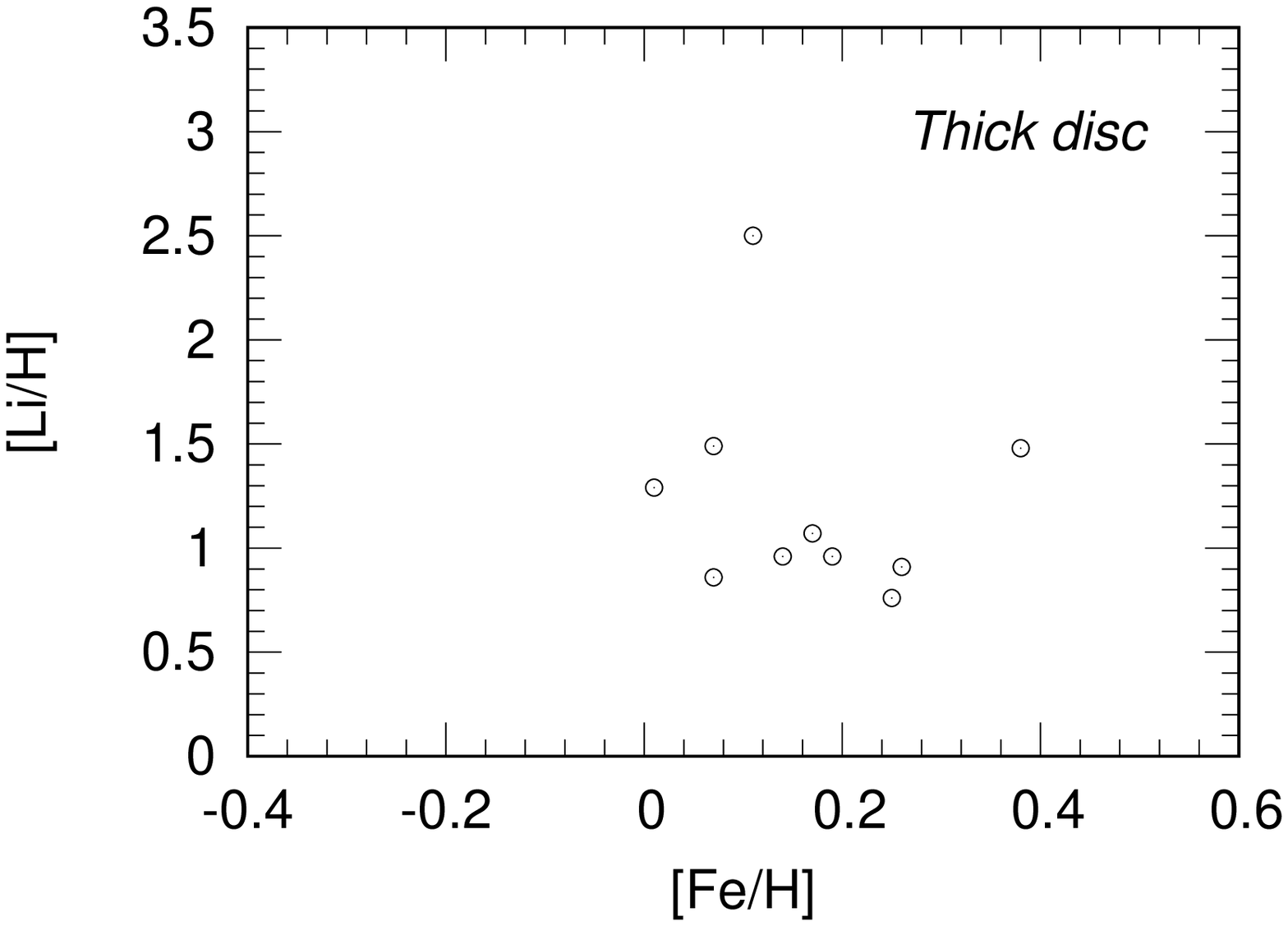}
\caption{Distribution of lithium abundances vs. [Fe/H] for our stars in the thin and thick discs. Empty and filled rings mark the non-SWP and SWP stars, respectively.
\label{_ttd}}
\end{figure*}

Majority of our stars are of the thin disc population, see fig. 1 in \cite{ivan17}. Nevertheless, we computed the distribution of lithium abundances for thick and thin discs, see Fig. \ref{_ttd}. Unfortunately, our sample of thick disc stars is not representative, but for the thin disc we see a rather uniform distribution of lithium abundances both in SWP and non-SWP stars of the thin disc. There are no discovered exoplanets around the thick disc stars from our sample. We do not find lithium rich stars (log N(Li)$>$1.5) at [Fe/H]$>$0.3, see Section \ref{_life}.

\subsection{log N(Li) vs. \Vm}

 We do not see any clear relation between microturbulence and Li abundance. Microturbulent motions in the atmospheres of stars of our sample characterise the properties of convective envelopes, which should be very similar for stars of our sample. We may suggest, at least qualitatively, that they are not connected with the global properties of the convective envelope which governs the processes of lithium burning at its lower boundary.

\section{\Li}
\label{_res1}

As we noted in the Introduction, the \Li isotopic ratio is of interest in many aspects of modern astrophysics. However, technically, the measurement of the \Li ratio is much more challenging, in comparison to the abundance determination. Indeed, the \liB and \liA resonance lines are very finely split, they form a common blend in the observed spectra, see Table \ref{_li67}. The \liA lines are located in the right wing of the \liB lines and in most cases the \liB lines are much stronger than the \liA. To analyse these lines, we should only use the highest quality observational data.

Before we can draw any conclusions, we should investigate two possible effects which can affect the \Li determination:

\subsection{Blending of the lithium resonance doublet lines by other elements.}

\begin{figure}
\begin{center}
\includegraphics [width=\linewidth, angle=0]{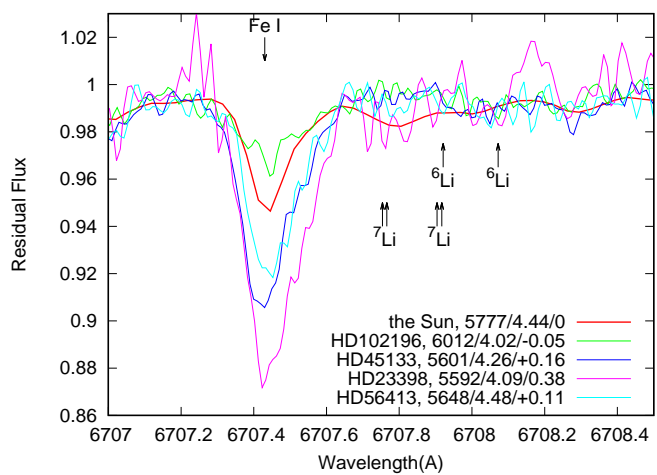}
\end{center}
\caption[]{Spectra of the five Li-poor stars and the Sun across the spectral range in which the Li resonance doublet lines form. \Teff, \logg and [Fe/H] are also shown, and the positions of the strongest \liB and \liA lines are marked by the arrows.}
\label{_noLi}
\end{figure}

In our case, we analysed the spectra of one quasi-uniform group of stars, whereby most of them are metal-rich dwarfs. In Fig. \ref{_noLi}, for some stars without detectable Li lines, we show the narrow spectral range in which these lines appear in spectra of Li-rich stars. We do not detect the lithium lines in these spectra behind the spuriosity of the spectral noise. We also show the solar spectrum from \cite{kuru84}, and the \liB lines are clearly seen in the solar spectrum. Furthermore, we see that the S/N in the solar spectrum is much higher than in our HARPS spectra.

\subsection{Precise fits to the observed Li lines}

Our HARPS spectra were observed with high enough resolution, i.e. R $\sim$ 100K, to permit accurate fitting of the lithium lines. However, not all of the stars show the presence of Li I lines. Therefore, we restricted our analysis to two SWP stars and two non-SWP stars,  that exhibited strong Li lines in their spectra. First of all, we note that in our case, the effective resolution is restricted by the effects of absorption lines broadening, not only by instrumental broadening and rotation, but also by the macroturbulent motions generated by the convective envelope.

The effects of macroturbulence have been well studied for the case of solar atmosphere, and it can be explained using more advanced, but very CPU consumable, 3D modelling, (see \citealp{mott17} and references therein). We work in the framework of 1D modelling methods, and to describe the general profile of the Li resonance blend formed by \liB+\liA lines,  we used  a more sophisticated approach in the treatment of the macroturbulent velocity distribution in comparison to the simple Gaussian used in the lithium abundance determination procedure. Indeed, \liA absorption affect the right wing of the stronger \liB line, see Fig. \ref{_noLi}. Therefore, to fit the wings of lines we should use a more advanced model of the macroturbulent velocity \vmac distribution, that allows to fit the full profile of the observed spectrum instead of a line core fitting analysis used in the abundance determination procedure.

Namely, for the procedure of \Li determination we used approach developed by \cite{smit76}. We adopted the broadening profile of the macroturbulent motions as Voigt functions. To avoid misunderstanding, we labeled the function as pseudo-Voigt $pV(x,a)$, where $x$ and $a$ are exponential and Lorentzian parameters of the convolving profile, respectively. In our case, the values $a$ and $x$ are determined by the macroturbulence and instrumental broadening profiles.

\begin{figure*}
\centering
\includegraphics[width=0.48\linewidth]{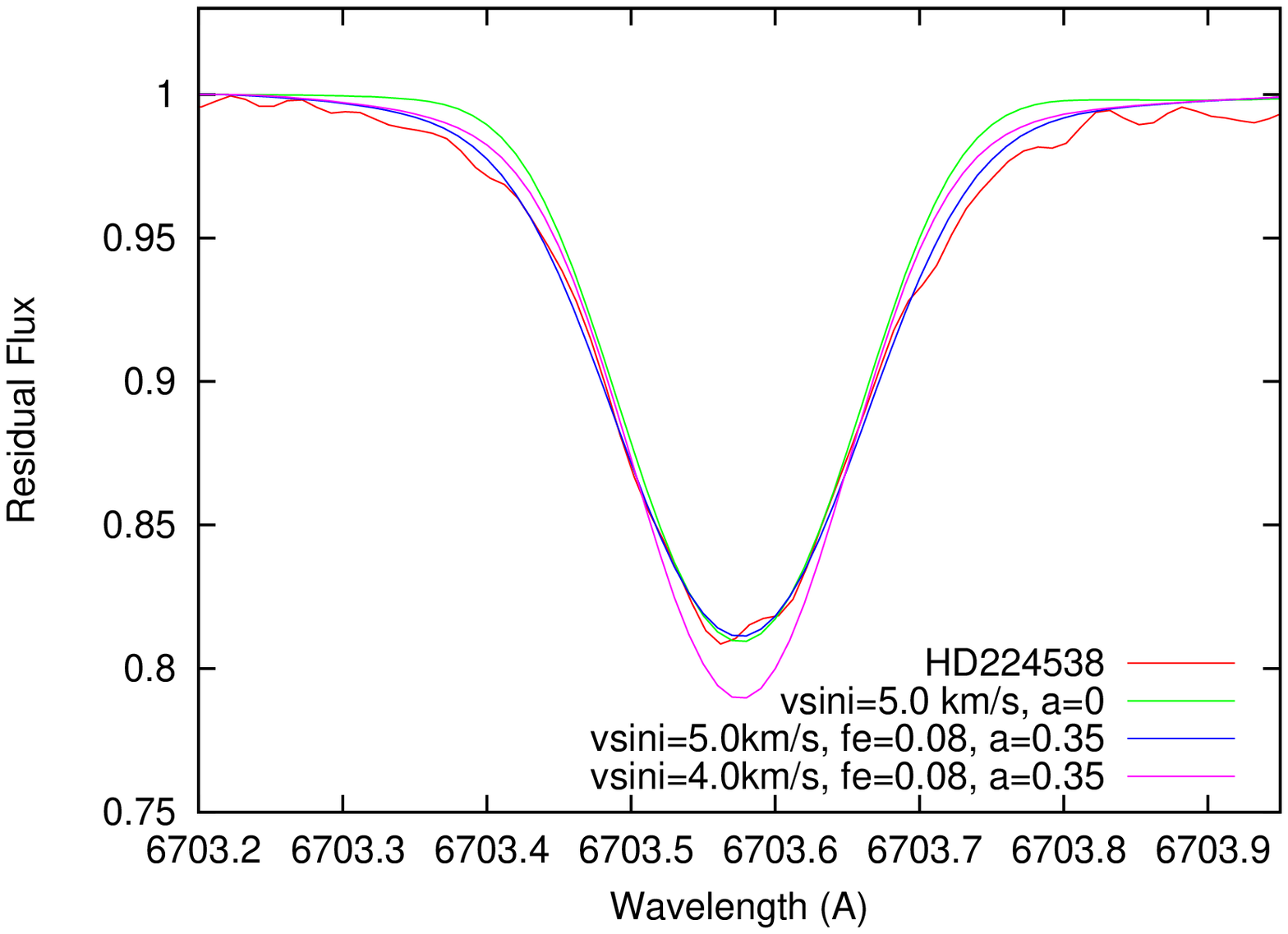}
\includegraphics[width=0.48\linewidth]{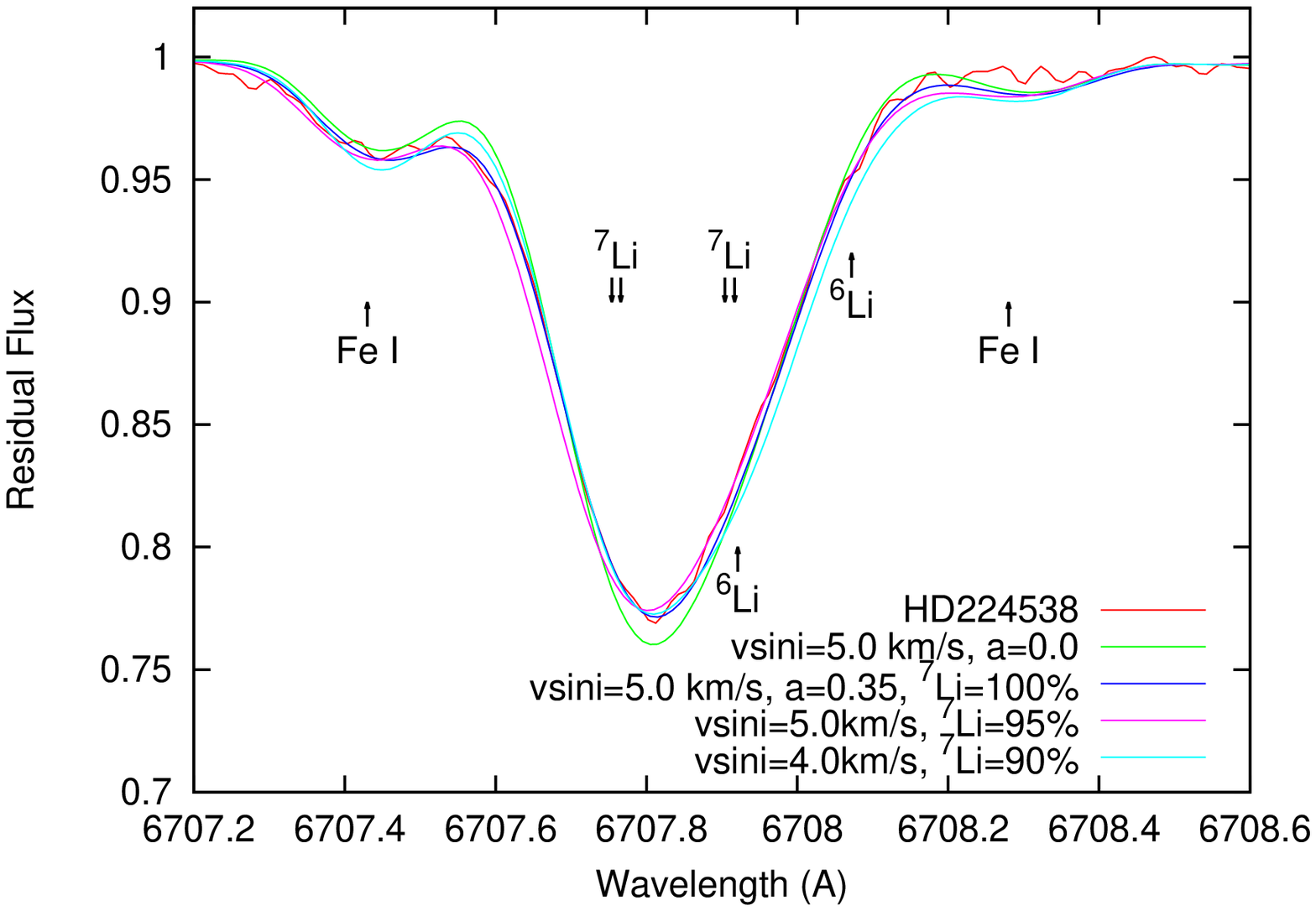}
\caption{{\it Left:} fits to the Fe I absorption line at 6703.58 \AA~ in the observed spectrum of the SWP star HD 224538 using the different parameters $a$ of our pseudo-Voigt function, 'Fe$=$0.08' marks changes of the Fe abundances due to use of the pseudo-Voigt broadening in comparison to a simple Gaussian. {\it Right}: fits to the observed Li resonance doublet with a different \Li. $^7Li$=90\% means here the computed spectra with $^7Li$=90\% and $^6Li$=10\%, i.e. \Li=9.}
\label{_224}
\end{figure*}

We followed the following scheme of \Li estimation:

\begin{itemize} 
\item From the fit to a single Fe I absorption line profile at 6703.58 \AA, we determined the parameter $a$ of the pseudo-Voigt function, describing the instrumental broadening and macroturbulence broadening, see left panel of the Fig. \ref{_224}. It is worth noting that implementing the 'pseudo-Voigt broadening' affects both the abundance and \vsini which we used to fit the observed profile of the absorption line.
\linebreak

\item Using the new parameters of the fits to the observed Fe I line, we provide the fit to observed Li I resonance doublets at 6708 \AA. In this case, to get the best fit to the Li line, we vary only the Li abundance and the \Li ratio. Li abundance variations do not exceed 0.05 dex, and the value is within the error bar of the Li abundance determination. Results are shown in the right panel of Fig. \ref{_224}.
\end{itemize}

The fits to the observed Li lines in the spectra of the other three stars that show notable Li lines, are provided in the Figs. \ref{_191}, \ref{_206} and \ref{_482} in the Appendix.

Unfortunately, our spectra are not of sufficiently high S/N to determine \Li ratio with high precision. Usually, the \Li isotopic ratio is determined from analysis of the spectra of much higher S/N$\sim$600 (see \citealp{mott17}). Still, for both of the stars we can determine the lower limit of \Li, i.e. \Li$>$10. From our fits we may conclude that, likely, \Li does not differ too much from the solar value, i.e. \Li=12.5 \citep{lang74}. To get more accurate estimations we should use spectra of higher quality, i.e. higher S/N. We plan to perform these studies in forthcoming works.

\section{Discussion}
\label{_dis}

In this paper we discuss the analysis of lithium abundances in the atmospheres of dwarfs and subgiants that are part of the CHEPS project, most of which are metal-rich. Therefore, the main conclusions to be drawn from this work regards the metal-rich population above the mean main sequence. If we make the supposition that the metallicity of stars generally increases with time, our sample of stars were formed in the post-solar epoch, therefore, lithium abundances in their atmospheres characterise the lithium formation processes at later epochs of our Galaxy's evolution.

We carried out our analysis using 1D model atmospheres, taking into account NLTE effects of the Li line formation. Generally speaking, NLTE computations within 3D model atmospheres exist, and may be more reliable. However, direct comparison of the 3D and 1D lithium abundance determination results shows that the differences do not exceed 0.02-0.04 dex for the case of red giants \citep{mott17}, (see discussions on LTE and NLTE in 1D and 3D atmospheres in \citealp{aspl03,mott17,klev16}).

In our case, the differences between 3D and 1D should be weaker due to the higher gravities of our stars. On the other hand, uncertainties of $\pm$50 K in the adopted effective temperatures provide the strongest effect, see Table \ref{_deli}. In our paper, we use the \Teff determined by photometric methods, meaning the application of multiple colour bands. Other input physical parameters, i.e. metallicities, gravities and rotational velocities were determined in an independent manner, see \cite{ivan17}, where a homogeneous set of the observed high resolution spectra was used. For our analysis, we only used the 6708 \AA~ lithium line due to heavy blending of the other lithium lines, particularly apparent in the case of our most metal-rich stars.

Our results of log N(Li) vs. \Teff appear to be similar to those obtained by other authors, see for example, fig. 4 in \cite{lope15}. We did not find any stars with notable lithium abundance inside the region of the so called 'lithium desert', located approximately at 5900$<$\Teff$<$6200 K, log N(Li)$<$2, in accordance with \cite{rami12}.  On the other hand, the number of stars with \Teff$>$5900 K with lithium (9) exceeds the number of stars with depleted lithium (4), see the left-top panel of Fig. \ref{_vs}.  The fact 
that at high \Teff stars have higher Li is explained naturally by the fact that they have thinner envelopes.

Lithium lines were found to be more abundant in the atmospheres of stars with a lower \logg$<$4, see top-right panel of Fig. \ref{_vs}. The ratio of stars with lithium to stars without detectable lithium is 8/3=2.6, which could reflect the fact that stars with lower \logg are younger. Lithium depletion as a function of the age of stars is discussed by \cite{baum10} and \cite{carl16}. On the other hand, we might expect to see the effects of increased shearing in a deeper convective envelope here. Therefore, more enhanced lithium depletion. Likely, we are observing a combination of both effects.

We do not find notable lithium lines in the spectra of stars with metallicities higher than 0.25 dex. A similar case can be seen in fig. 5 of \cite{rami12}. Although, the primary effect on the lithium abundance is the time dependent lithium depletion, which depends on the age of star. A high opacity of the metal-rich stellar atmospheres results in an increased convection, which then leads to an increased depletion of lithium. On the other hand, lithium is less depleted in the atmospheres of fast rotators (see the middle right panel of Fig. \ref{_vs}).  This was also observed in the Pleiades by \cite{sode93}.

Hypothetically, we may suggest that these stars formed in the local giant molecular cloud that has been photo-evaporated over long timescale by nearby OB associations that went supernovae, seeding the environment with the metals we see in these stars, but leaving the remaining gas depleted in lithium.

It appears that we have higher chance to detect planets near slowly rotating stars. About 10\% of our targets are known to host planets. Most of the known planets in our CHEPS sample were discovered near comparatively slow rotators with \vsini$<$3 \kmps, in accordance with the other authors, (e.g. \citealp{take10}). In general, this agrees  as well with the work of \cite{gonz10a}, that the SWP have smaller \vsini values than the stars without detected planets. However, this also  would be the result of observational bias. Rapid rotators show broader atomic lines, complicating the measurement of small Doppler shits in the radial velocity measurements, e.g. \citealp{jenk13}. In our case, we obtained the well known correlation for the stars of different ages between the lithium abundance and \vsini (\citealp{mart94,mess16,beck17}). However, for the older stars, observed lithium abundances show spread that indicate the dependence on their ages and masses \citep{jone97,sode93}. Most of our stars are field stars. However, they are on the main sequence, or just leaving. Therefore, we still see the well pronounced dependence of log N(Li) vs. \vsini.  
 It is clear that the general trend shows higher abundances of Li
for faster rotators, except for a few fast rotators that show low Li,
particularly the ''strange'' planet host HD 147973 that rotates at
8km/s yet shows no Li. This is a significant difference given 
that it is also a hot star \Teff = 5972 K with low \logg =3.90 and a
[Fe/H] =-0.1, meaning its low Li might not be expected at all. It is
likely that HD 147973 is low mass star, i.e. a star with a thick
convective envelope, which is on the way to the main sequence. On the
other hand, when we look at the stars with rotational velocities
around 3 km/s, all the planet hosts have low Li, and it is clear that
they are not cooler than the stars with similar rotational velocities
that are non-planet hosts, at least in a statistical sense. 
Our analysis does not show any significant differences in
the lithium abundance between planet host stars and those without known planets.

Most of our stars are located at distances between 60-120 pc, due to the selection bias present in the original CHEPS sample that was made to select against stars currently on any other survey. We have only two stars that show lithium at d$<$50 pc, and 10 stars without lithium.

The majority of our stars show \Vm$>$1.1 \kmps and it is a region of the parameter space where we see most of the stars with strong lithium, along with the most of the SWP as well, see the bottom right panel of Fig. \ref{_vs}.

In our sample of 107 stars, measurable lithium lines were found in 43 targets without planets and 2 SWP. Only upper limits of the lithium abundance were determined for 54 stars without planets and 8 SWP. The ratio of stars with lithium to stars without lithium consists of 43/54=80\%. In total, we obtained a ratio of the known SWP with strong lithium to the stars without lithium of 2/8=25\%,  i.e. we found a significantly lower proportion of stars with detectable Li among known planet hosts than among stars without planets.

On the other hand, all our SWP with  the measurable Li abundance follow the known trend log N(Li) vs. \vsini for the 'Li-rich' stars, see \cite{sode93} and Section \ref{_vsinili}. This is likely a signature of the SWP and should motivate a continued and expanded search among the CHEPS stars with log N(Li)$>$2. Unfortunately, the current sample is not large enough to provide precisely constrained results for the presence of lithium in SWP compared to those without. Therefore, we plan to carry out an expanded investigation along these lines that cover a larger sample of stars.

In Fig. \ref{_noLi} we show the comparison of spectra of our stars with the spectrum of the Sun, as a star, by \cite{kuru84}. The lines of lithium in the solar spectrum are weak, where the equivalent widths of the lithium resonance doublet lines are 1.8 and 0.8 $\mu$\AA~ at the center of the solar disc \citep{brau75}. In fact,  the Li I lines intensity in the solar spectrum is comparable to the intensities of the weakest stellar lines, used in this paper to determine the upper limit of the lithium abundances due to lower quality of our spectra. Nevertheless, lithium in the Sun follows the general trends seen in Fig.~\ref{_vs}. At the very least, low lithium abundance in the Sun agrees with the results for the other stars of the same metallicity, \Teff, \logg, \vsini and \Vm.

Determination of \Li would improve our knowledge of the stars in our sample. Unfortunately, our spectra are not of a sufficient S/N to determine \Li with high precision. For our two SWP and two non-SWP stars we determined only the lower limit of the \Li, i.e. \Li$>$10. Likely, their \Li does not differ too much from the solar value. We plan to get more accurate \Li estimations for the SWP and non-SWP stars using fits to a higher quality spectra, i.e. higher S/N.

In summary, our analyses confirm the problems faced when trying to study possible connection between the presence of stellar atmospheric lithium and giant planets orbiting stars. Nowadays, we witness the results of various processes that are all at play at the same time, processes that are variable in time and space and relate to the sink/formation of lithium atoms in different parts of Galaxy, see Section \ref{_intro}.  It is likely that lithium abundance evolution of the stars on the main sequence changes with time and might be modulated by the long period magnetic activity cycles, if substantial amount of lithium could be produced by spallation reactions during strong flares, see \cite{wall69} and Section \ref{_forsink}. Finally, stars form and evolve in the variable (in time and space) interstellar environment. It looks like the lithium-planet hosting connection is of a secondary importance in stellar evolution, if it does exist at all. Still, it may be more significant across different evolutionary epochs. To establish the true picture, we should continue to carry out the fine analysis of homogeneously observed and analysed extended samples that consist of the SWP and the stars without planets, particularly focused on well constrained evolutionary phases.

\section{Acknowledgements}

We acknowledge funding by EU PF7 Marie Curie Initial Training Networks (ITN) RoPACS project (GA N 213646) and the special support by the NAS Ukraine under the Main Astronomical Observatory GRAPE/GPU/GRID computing cluster project. We also acknowledge support by Fondecyt grant 1161218 and partial support by CATA-Basal (PB06, CONICYT) and support from the UK STFC via grants ST/M001008/1 and Leverhulme Trust RPG-2014-281.

We thank the anonymous Referee for his/her thorough review and highly appreciate the comments and suggestions, which significantly contributed to improving the quality of the publication.

\bibliographystyle{aa}
\bibliography{mnemonic,biblio_Li}

\begin{onecolumn}
\begin{appendix}
\section {A}

\begin{figure*}
\centering
\includegraphics[width=0.48\linewidth]{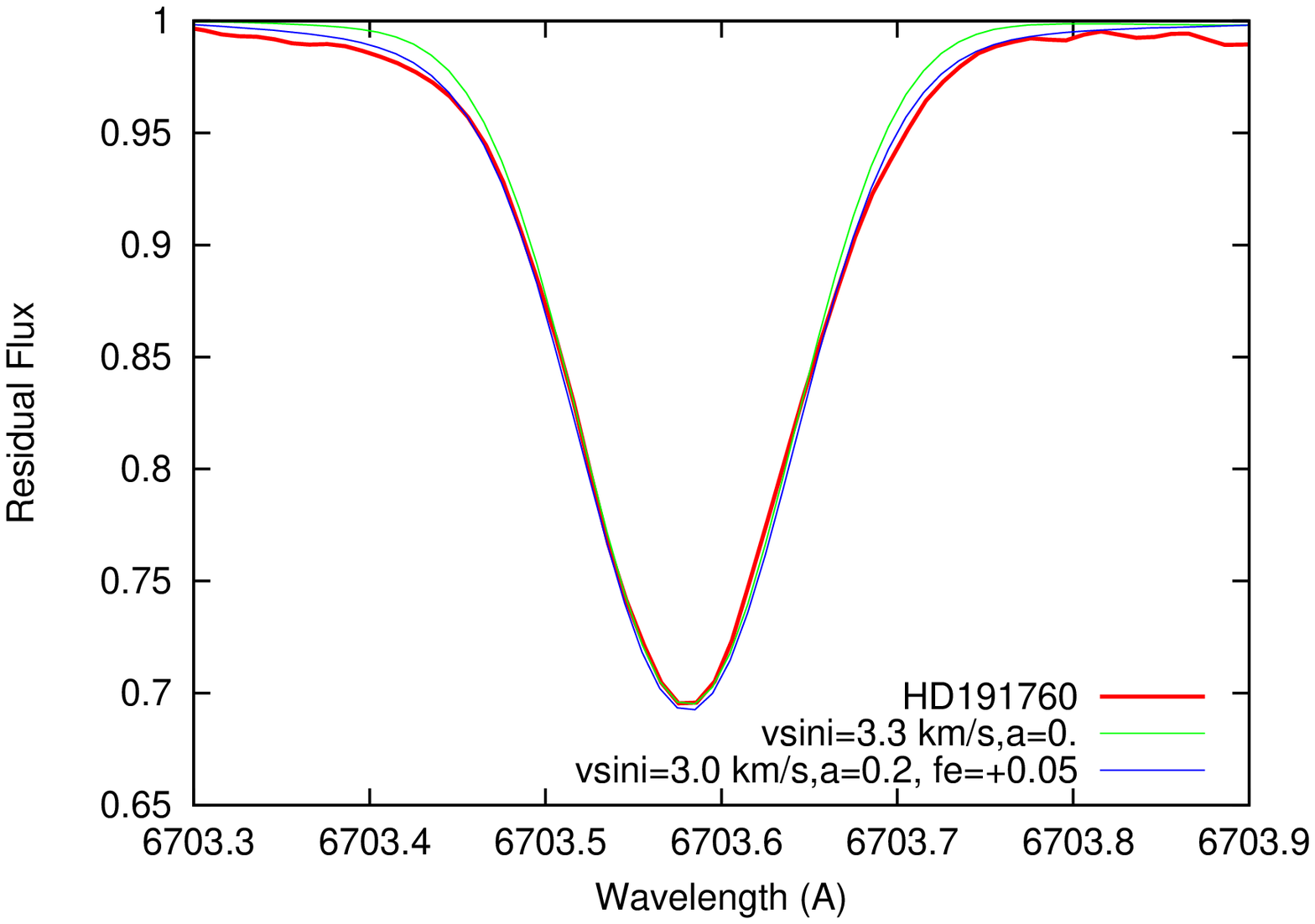}
\includegraphics[width=0.48\linewidth]{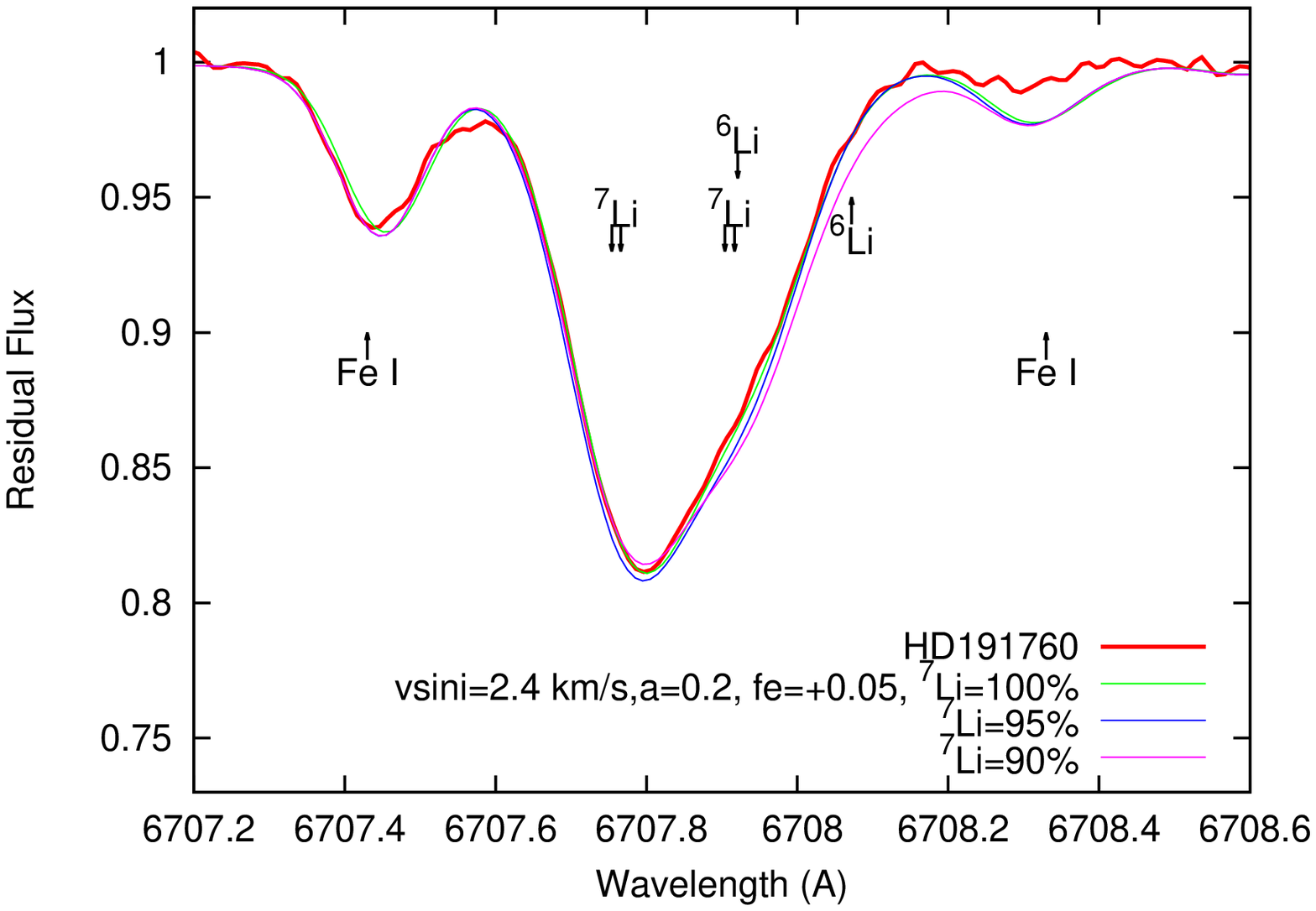}
\caption{{\it Left:} The fits to the Fe I absorption line at 6703.58 \AA~ in the observed HD 191760 spectrum using the different parameters $a$ of our pseudo-Voigt function. {\it Right}: Fits to observed Li resonance doublet with different \Li.}
\label{_191}
\end{figure*}

\begin{figure*}
\centering
\includegraphics[width=0.48\linewidth]{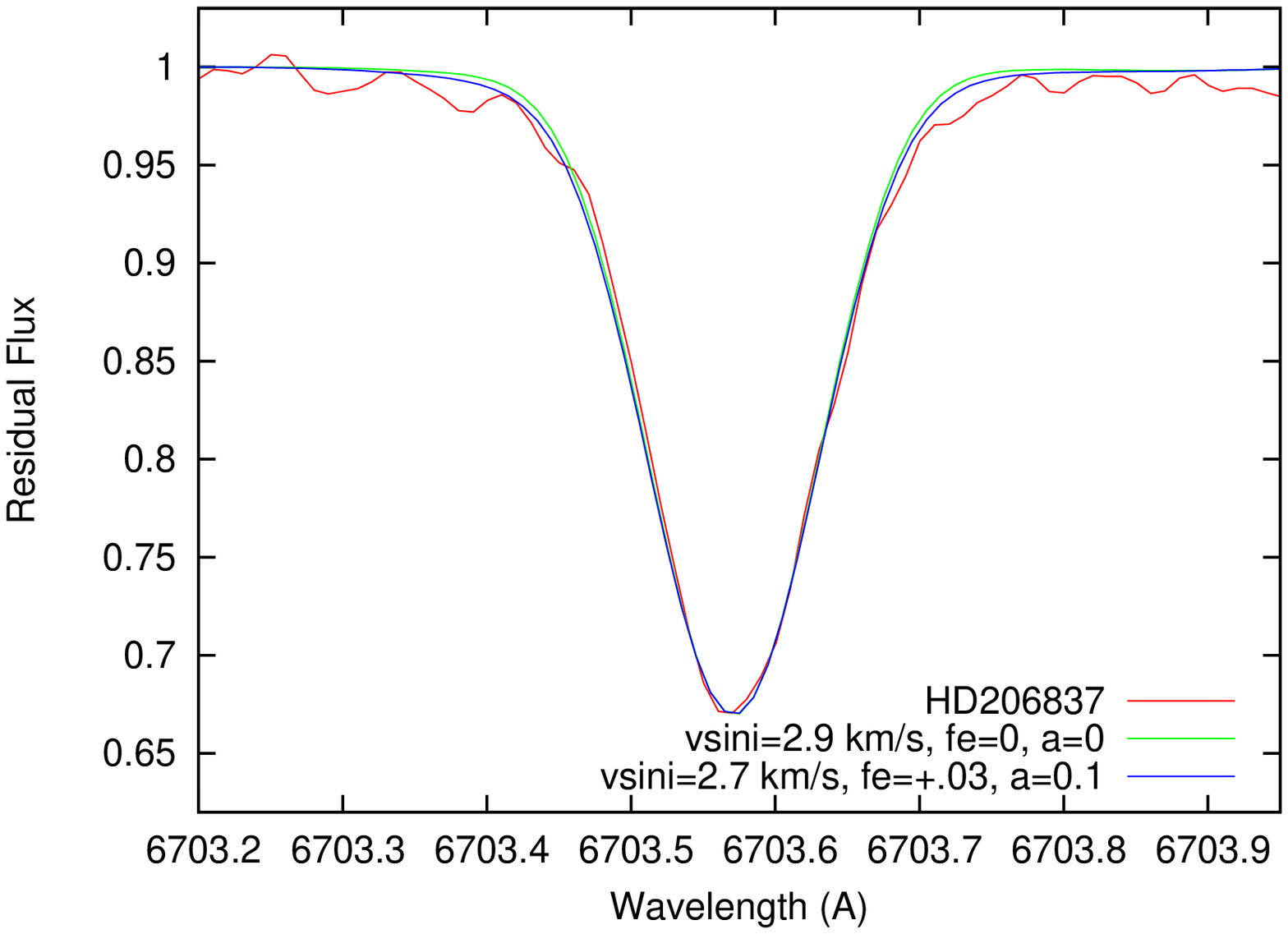}
\includegraphics[width=0.48\linewidth]{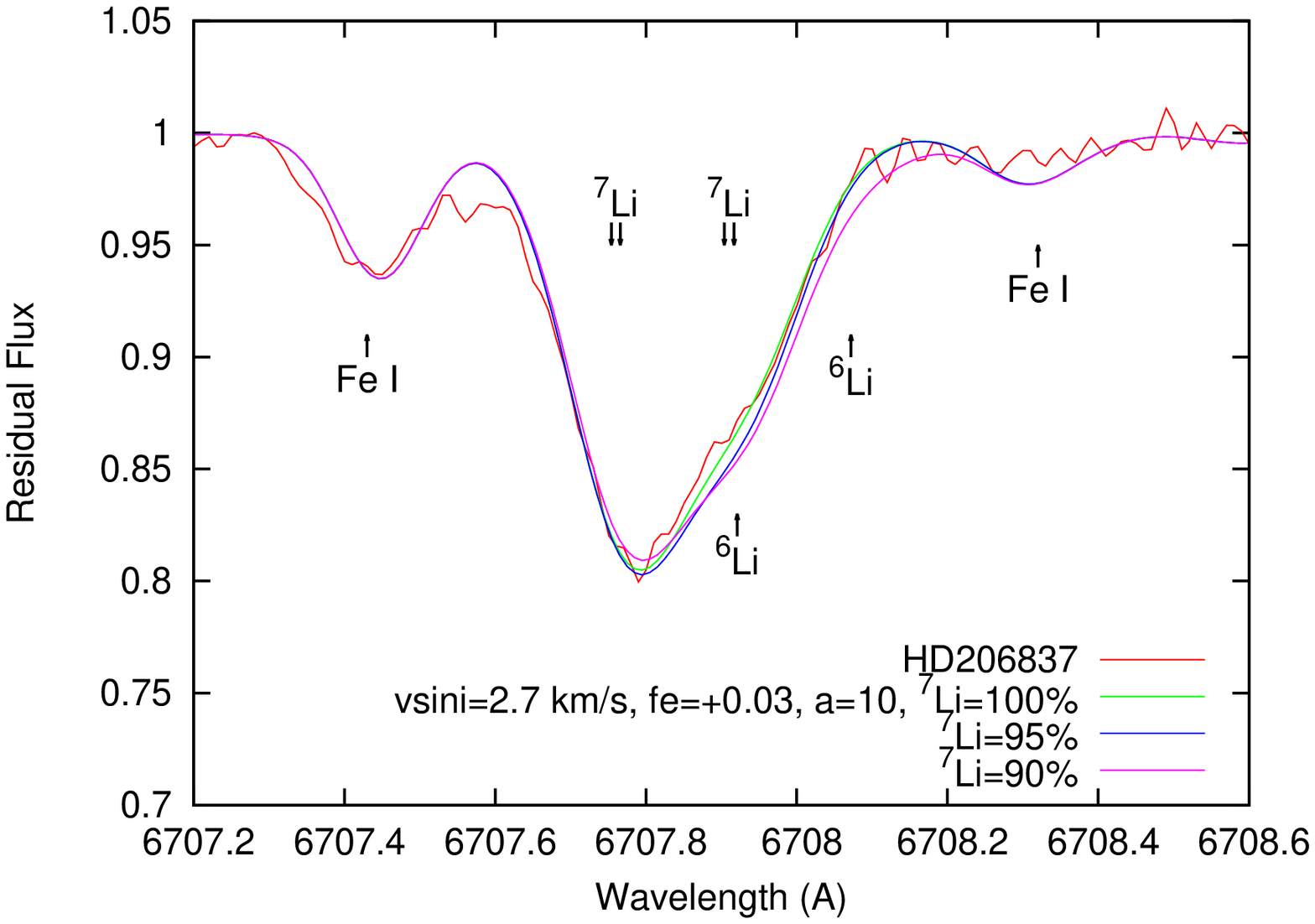}
\caption{{\it Left:} The fits to the Fe I absorption line at 6703.58 \AA~ in the observed HD 206837 spectrum using the different parameters $a$ of our pseudo-Voigt function. {\it Right}: Fits to observed Li resonance doublet with different \Li.}
\label{_206}
\end{figure*}

\begin{figure*}
\centering
\includegraphics[width=0.48\linewidth]{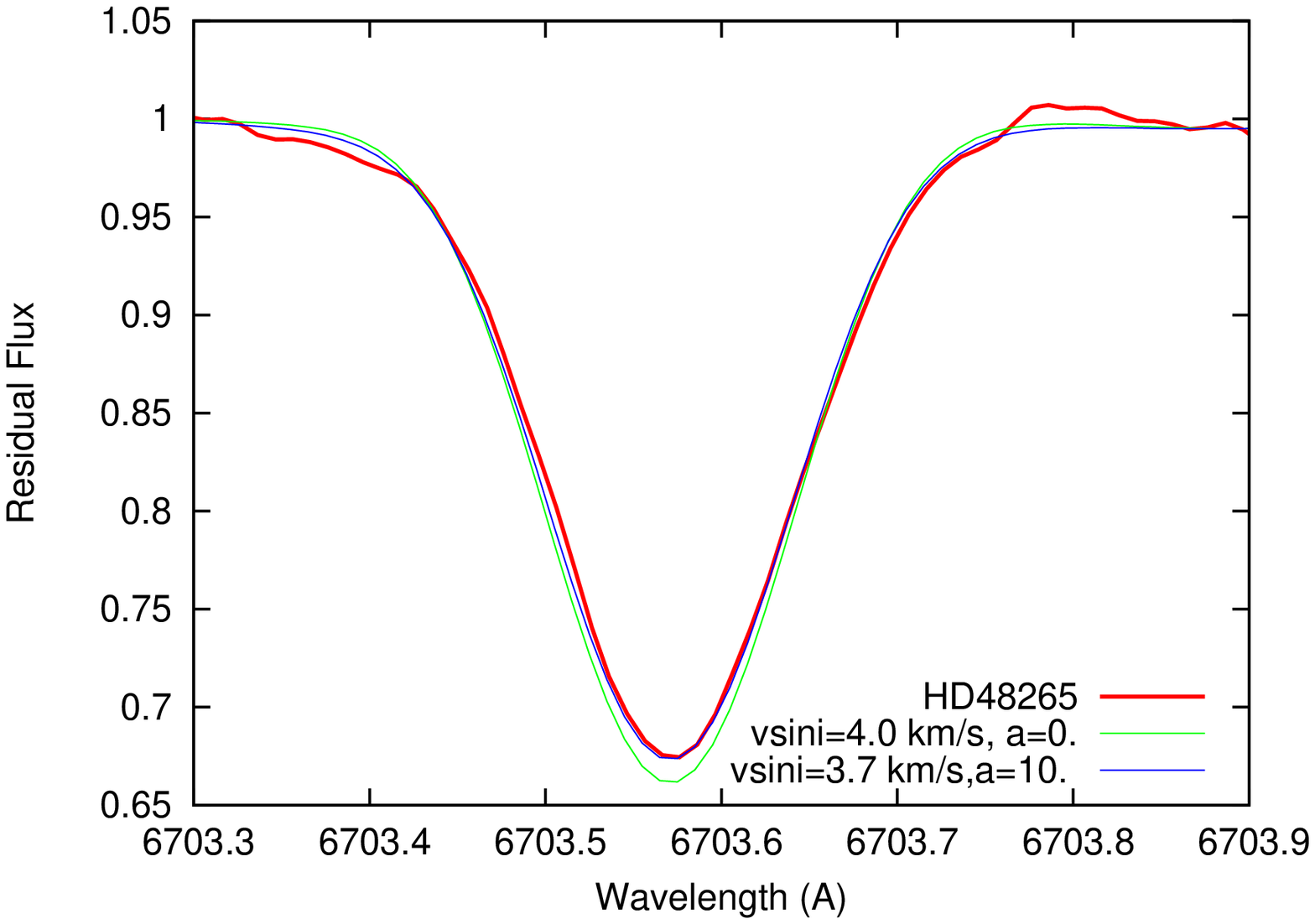}
\includegraphics[width=0.48\linewidth]{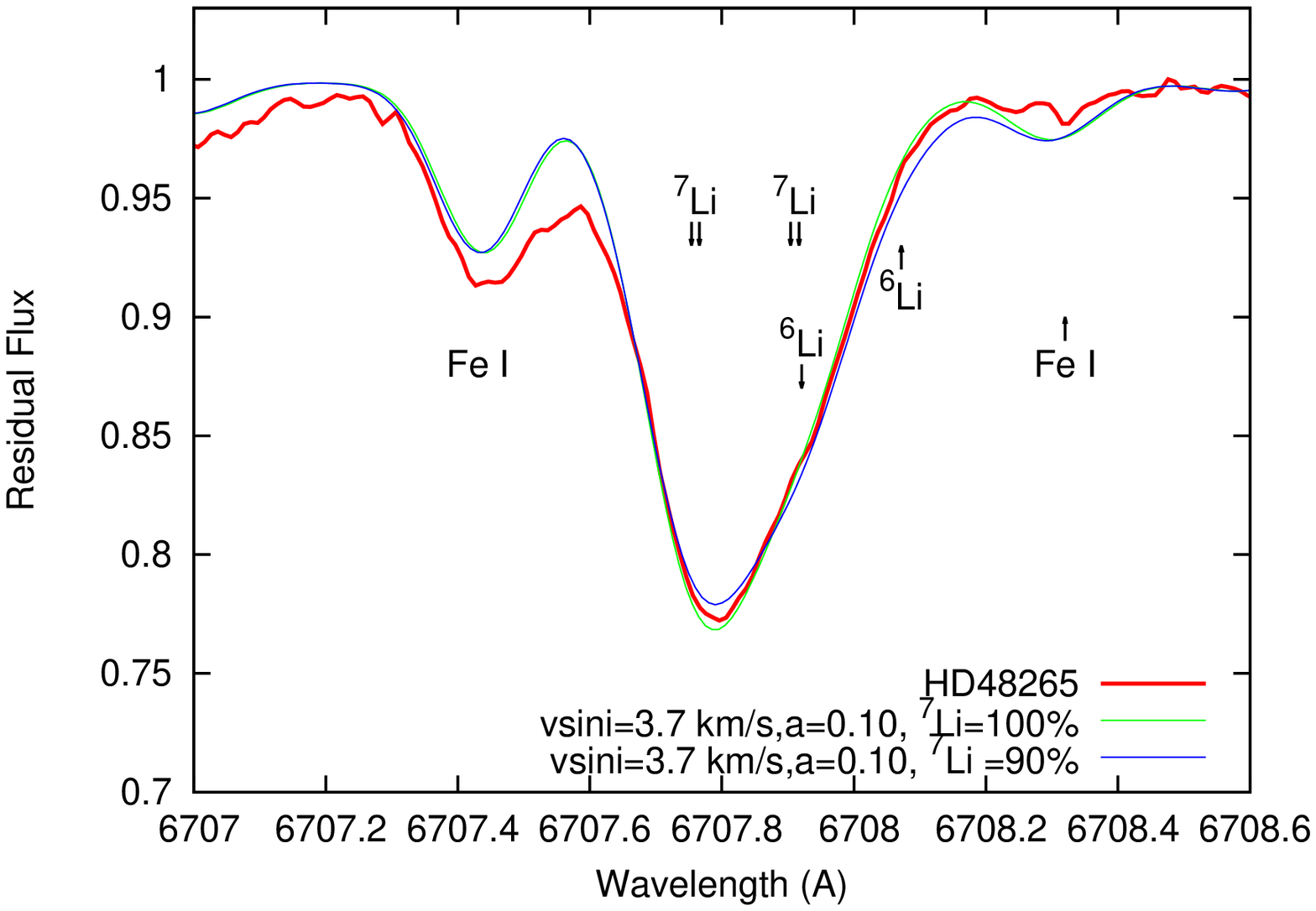}
\caption{{\it Left:} The fits to the Fe I absorption line at 6703.58 \AA~ in the observed SWP star HD48265 spectrum using the different parameters $a$ of our pseudo-Voigt function. {\it Right}: Fits to observed Li resonance doublet with different \Li.}
\label{_482}
\end{figure*}

\begin{table}
\caption{Lithium line parameters used in this work \citep{kuru95}.}
\label{_li67}
\begin{tabular}{cccccc}
\hline\hline

&&&&& \\
$\lambda(A)$    & $gf$    &  $E''$    &  $c_2$ & $c_4$ & $c_6$ \\
&&&&& \\
\hline
&&&&& \\
\liB &&&&& \\
 6707.756 &3.733E-01 & 0.00 & 7.56 &-5.78 &-7.57  \\
 6707.768 &6.223E-01 & 0.00 & 7.56 &-5.78 &-7.57  \\ 
 6707.907 &1.556E-01 & 0.00 & 7.56 &-5.78 &-7.57  \\ 
 6707.908 &3.112E-02 & 0.00 & 7.56 &-5.78 &-7.57  \\
 6707.919 &1.556E-01 & 0.00 & 7.56 &-5.78 &-7.57  \\ 
 6707.920 &1.556E-01 & 0.00 & 7.56 &-5.78 &-7.57  \\ 
\liA &&&&&                                        \\ 
 6707.920 &3.319E-01 & 0.00 & 7.56 &-5.78 &-7.57  \\
 6707.923 &6.637E-01 & 0.00 & 7.56 &-5.78 &-7.57  \\ 
 6708.069 &1.476E-01 & 0.00 & 7.56 &-5.78 &-7.57  \\ 
 6708.070 &1.845E-02 & 0.00 & 7.56 &-5.78 &-7.57  \\ 
 6708.074 &1.845E-01 & 0.00 & 7.56 &-5.78 &-7.57  \\
 6708.075 &1.476E-01 & 0.00 & 7.56 &-5.78 &-7.57  \\
\end{tabular}
\end{table}

\begin{table*}                    
\caption{SWP with measurable abundance of lithium and with upper limit of lithium in their atmospheres
\label{_lires11}}
\begin{tabular}{cccc cccc cccc}
\hline\hline
\multicolumn{12}{c}{SWP with measurable lithium} \\
Name        & \NLTE  &  $d$ (pc) &  \Teff &  \logg   &     \Vm      &    \vsini     &       [Fe]      & $M_j$  & $P(d)$ &  $a(AU)$ & $e$ \\
HD 48265    &  2.47  &   85.40   &  5651  &   3.92   & 1.4$\pm$ 0.2 &  4.0$\pm$ 0.1 &  0.17$\pm$ 0.01 &  1.16  &   700  &  1.51  &  0.18 \\
HD 224538   &  2.75  &   77.76   &  6097  &   4.29   & 1.4$\pm$ 0.2 &  5.1$\pm$ 0.1 &  0.09$\pm$ 0.02 &  5.97  &  1189  &  2.28  &  0.46 \\ 
            &&&& &&&& &&& \\ \hline
            \multicolumn{12}{c}{SWP with upper limit of lithium} \\ 
Name        & \NLTE  & $d$ (pc)  &  \Teff &  \logg   &     \Vm      &    \vsini     &       [Fe]      & $M_j$  & $P(d)$ &  $a(AU)$ & $e$ \\
HD 128356   &  0.61  &   26.03   &  4875  &   4.58   & 0.8$\pm$ 0.2 &  1.3$\pm$ 0.1 &  0.34$\pm$ 0.03 &  0.89  &   298  &  0.87  &  0.57 \\
HD 147873   &  0.63  &  104.93   &  5972  &   3.90   & 1.4$\pm$ 0.2 &  8.3$\pm$ 0.2 & -0.09$\pm$ 0.02 &  5.14  &   117  &  0.52  &  0.20 \\
HD 77338    &  0.80  &   40.75   &  5315  &   4.42   & 1.2$\pm$ 0.2 &  1.3$\pm$ 0.1 &  0.16$\pm$ 0.02 &  0.50  &     6  &  0.06  &   -   \\ 
HD 143361   &  0.95  &   65.66   &  5505  &   4.42   & 1.0$\pm$ 0.2 &  1.7$\pm$ 0.1 &  0.18$\pm$ 0.01 &  3.12  &  1057  &  2.00  &  0.15 \\
HD 165155   &  1.07  &   64.98   &  5426  &   4.57   & 1.0$\pm$ 0.2 &  1.7$\pm$ 0.1 &  0.07$\pm$ 0.02 &  2.89  &   435  &  1.13  &  0.20 \\
HD 154672   &  1.12  &   64.77   &  5655  &   4.16   & 1.2$\pm$ 0.2 &  3.0$\pm$ 0.1 &  0.10$\pm$ 0.02 &  5.02  &   164  &  0.60  &  0.61 \\
HD 152079   &  1.29  &   83.33   &  5726  &   4.35   & 1.2$\pm$ 0.2 &  2.6$\pm$ 0.1 &  0.16$\pm$ 0.03 &  3.00  &  2097  &  3.20  &  0.60 \\
HD 9174     &  1.44  &   78.93   &  5577  &   4.05   & 1.2$\pm$ 0.2 &  2.9$\pm$ 0.1 &  0.26$\pm$ 0.01 &  1.11  &  1179  &  2.20  &  0.12 \\            
\end{tabular}
\end{table*}

\begin{table*}                    
\caption{Measurable abundances of lithium in the atmospheres of stars without detected planets
\label{_lires10}}
\begin{tabular}{cccccccccc}
\hline\hline
Name       &   Distance &  \Teff &  \logg  &     \Vm       &     \vsini    &       Fe I       &  \NLTE &   \LTE  &  \dNLTE \\  
HD 6790    &   105.93   &  6012  &   4.40  &  0.8$\pm$ 0.2 &  4.7$\pm$ 0.2 & -0.06$\pm$ 0.02  &   2.51 &   2.47  &   0.04  \\ 
HD 7950    &   117.37   &  5426  &   3.94  &  1.2$\pm$ 0.2 &  2.7$\pm$ 0.1 &  0.11$\pm$ 0.02  &   1.65 &   1.54  &   0.11  \\ 
HD 8446    &    73.64   &  5819  &   4.14  &  1.2$\pm$ 0.2 &  3.9$\pm$ 0.1 &  0.13$\pm$ 0.02  &   2.70 &   2.66  &   0.04  \\ 
HD 10188   &   107.64   &  5714  &   4.16  &  1.2$\pm$ 0.2 &  3.8$\pm$ 0.1 &  0.18$\pm$ 0.02  &   2.79 &   2.75  &   0.04  \\ 
HD 10278   &    56.31   &  5712  &   4.62  &  1.0$\pm$ 0.2 &  2.9$\pm$ 0.1 &  0.01$\pm$ 0.02  &   1.67 &   1.60  &   0.07  \\ 
HD 18708   &    62.85   &  5838  &   4.36  &  1.2$\pm$ 0.2 &  3.6$\pm$ 0.1 &  0.03$\pm$ 0.01  &   1.89 &   1.82  &   0.07  \\ 
HD 18754   &    97.66   &  5531  &   3.84  &  1.4$\pm$ 0.2 &  3.2$\pm$ 0.1 &  0.03$\pm$ 0.01  &   2.35 &   2.26  &   0.09  \\ 
HD 19493   &    75.76   &  5743  &   4.16  &  1.2$\pm$ 0.2 &  3.0$\pm$ 0.1 &  0.12$\pm$ 0.01  &   2.03 &   1.96  &   0.07  \\ 
HD 19773   &    65.36   &  6156  &   4.13  &  1.2$\pm$ 0.2 &  4.2$\pm$ 0.1 &  0.03$\pm$ 0.02  &   2.91 &   2.90  &   0.01  \\ 
HD 38467   &    68.35   &  5721  &   4.18  &  1.2$\pm$ 0.2 &  2.8$\pm$ 0.1 &  0.10$\pm$ 0.02  &   1.97 &   1.89  &   0.08  \\ 
HD 42538   &    87.95   &  5939  &   3.98  &  1.4$\pm$ 0.2 &  5.7$\pm$ 0.1 & -0.04$\pm$ 0.01  &   2.61 &   2.57  &   0.04  \\ 
HD 55524   &   109.41   &  5700  &   4.22  &  1.2$\pm$ 0.2 &  3.5$\pm$ 0.1 &  0.14$\pm$ 0.02  &   2.21 &   2.13  &   0.08  \\ 
HD 56957   &    61.35   &  5674  &   4.09  &  1.4$\pm$ 0.2 &  3.5$\pm$ 0.1 &  0.14$\pm$ 0.02  &   2.25 &   2.17  &   0.08  \\ 
HD 66653   &    37.13   &  5771  &   4.42  &  1.2$\pm$ 0.2 &  3.2$\pm$ 0.1 & -0.05$\pm$ 0.02  &   1.95 &   1.88  &   0.07  \\ 
HD 78130   &    60.57   &  5744  &   4.43  &  1.0$\pm$ 0.2 &  2.9$\pm$ 0.1 &  0.04$\pm$ 0.02  &   1.97 &   1.90  &   0.07  \\ 
HD 86006   &    76.39   &  5668  &   3.97  &  1.2$\pm$ 0.2 &  3.6$\pm$ 0.1 &  0.15$\pm$ 0.02  &   1.92 &   1.83  &   0.09  \\ 
HD 90028   &    84.10   &  5740  &   4.06  &  1.2$\pm$ 0.2 &  4.2$\pm$ 0.2 &  0.09$\pm$ 0.02  &   2.54 &   2.48  &   0.06  \\ 
HD 90520   &    66.53   &  5870  &   4.08  &  1.2$\pm$ 0.2 &  5.0$\pm$ 0.1 &  0.06$\pm$ 0.02  &   2.55 &   2.50  &   0.05  \\ 
HD 101197  &    82.99   &  5756  &   4.23  &  1.0$\pm$ 0.2 &  4.5$\pm$ 0.2 &  0.04$\pm$ 0.02  &   2.20 &   2.13  &   0.07  \\ 
HD 101348  &    81.43   &  5620  &   3.95  &  1.4$\pm$ 0.2 &  4.1$\pm$ 0.1 &  0.11$\pm$ 0.02  &   2.50 &   2.43  &   0.07  \\ 
HD 102361  &    83.61   &  5978  &   4.12  &  1.4$\pm$ 0.2 &  7.4$\pm$ 0.3 & -0.15$\pm$ 0.02  &   2.72 &   2.69  &   0.03  \\ 
HD 105750  &   118.34   &  5672  &   4.24  &  0.8$\pm$ 0.2 &  4.6$\pm$ 0.1 &  0.06$\pm$ 0.02  &   2.37 &   2.30  &   0.07  \\ 
HD 107181  &    85.18   &  5581  &   4.17  &  1.2$\pm$ 0.2 &  2.8$\pm$ 0.1 &  0.22$\pm$ 0.02  &   1.91 &   1.81  &   0.10  \\ 
HD 127423  &    68.31   &  6020  &   4.26  &  1.0$\pm$ 0.2 &  4.3$\pm$ 0.1 & -0.09$\pm$ 0.02  &   2.72 &   2.70  &   0.02  \\ 
HD 143120  &    67.29   &  5576  &   3.95  &  1.2$\pm$ 0.2 &  3.7$\pm$ 0.1 &  0.16$\pm$ 0.02  &   2.34 &   2.25  &   0.09  \\ 
HD 144550  &    86.06   &  5652  &   4.19  &  1.2$\pm$ 0.2 &  3.8$\pm$ 0.1 &  0.08$\pm$ 0.02  &   2.35 &   2.27  &   0.08  \\ 
HD 144848  &    78.93   &  5777  &   4.26  &  1.2$\pm$ 0.2 &  3.5$\pm$ 0.1 &  0.10$\pm$ 0.01  &   1.86 &   1.78  &   0.08  \\ 
HD 144899  &   130.04   &  5833  &   4.13  &  1.2$\pm$ 0.2 &  3.5$\pm$ 0.1 &  0.17$\pm$ 0.02  &   2.69 &   2.65  &   0.04  \\ 
HD 149189  &    67.16   &  5771  &   4.08  &  1.4$\pm$ 0.2 &  3.6$\pm$ 0.1 &  0.04$\pm$ 0.01  &   2.34 &   2.28  &   0.06  \\ 
HD 154221  &    61.01   &  5797  &   4.47  &  1.0$\pm$ 0.2 &  3.0$\pm$ 0.1 &  0.01$\pm$ 0.02  &   1.78 &   1.71  &   0.07  \\ 
HD 158469  &    72.15   &  6105  &   4.19  &  1.2$\pm$ 0.2 &  5.3$\pm$ 0.1 & -0.14$\pm$ 0.01  &   2.58 &   2.55  &   0.03  \\ 
HD 189627  &    70.92   &  6210  &   4.40  &  1.4$\pm$ 0.2 &  7.8$\pm$ 0.1 &  0.07$\pm$ 0.02  &   3.01 &   3.01  &   0.00  \\ 
HD 190125  &    86.96   &  5644  &   4.53  &  1.0$\pm$ 0.2 &  3.3$\pm$ 0.1 &  0.04$\pm$ 0.02  &   1.71 &   1.63  &   0.08  \\ 
HD 191122  &    69.16   &  5851  &   4.34  &  1.2$\pm$ 0.2 &  3.0$\pm$ 0.2 &  0.10$\pm$ 0.02  &   1.90 &   1.83  &   0.07  \\ 
HD 191760  &    81.63   &  5816  &   4.10  &  1.4$\pm$ 0.2 &  2.9$\pm$ 0.1 &  0.07$\pm$ 0.01  &   2.42 &   2.36  &   0.06  \\ 
HD 193995  &    92.34   &  5661  &   4.09  &  1.2$\pm$ 0.2 &  3.2$\pm$ 0.1 &  0.11$\pm$ 0.02  &   1.58 &   1.50  &   0.08  \\ 
HD 194490  &    72.99   &  5854  &   4.44  &  1.0$\pm$ 0.2 &  3.3$\pm$ 0.1 & -0.04$\pm$ 0.02  &   1.76 &   1.70  &   0.06  \\ 
HD 206683  &    67.84   &  5909  &   4.37  &  1.2$\pm$ 0.2 &  3.8$\pm$ 0.1 &  0.14$\pm$ 0.02  &   2.25 &   2.19  &   0.06  \\ 
HD 206837  &   166.94   &  5616  &   4.07  &  1.2$\pm$ 0.2 &  2.9$\pm$ 0.1 & -0.01$\pm$ 0.02  &   2.26 &   2.18  &   0.08  \\ 
HD 221954  &    99.01   &  5602  &   4.10  &  1.2$\pm$ 0.2 &  2.8$\pm$ 0.1 &  0.19$\pm$ 0.02  &   1.57 &   1.48  &   0.09  \\ 
HD 222910  &   122.55   &  5480  &   4.05  &  1.2$\pm$ 0.2 &  2.4$\pm$ 0.1 &  0.13$\pm$ 0.02  &   1.54 &   1.43  &   0.11  \\ 
HIP 19807  &   106.84   &  5892  &   4.54  &  1.0$\pm$ 0.2 &  3.0$\pm$ 0.1 &  0.07$\pm$ 0.01  &   1.82 &   1.76  &   0.06  \\ 
HIP 31831  &   107.76   &  5845  &   4.29  &  1.2$\pm$ 0.2 &  3.6$\pm$ 0.1 &  0.16$\pm$ 0.01  &   2.41 &   2.35  &   0.06  \\ 
\end{tabular}
\end{table*}

\begin{table*}                    
\caption{Upper limits of the lithium abundances in the atmospheres of stars without detected planets
\label{_lires00}}
\begin{tabular}{cccc cccc cc}
\hline\hline
Name       &   Distance & \Teff  &  \logg &     \Vm      &     \vsini    &       [Fe]      &   \NLTE  &   \LTE   &  \dNLTE  \\ 
\hline
HD 8389    &    30.51   &  5243  &   4.52 & 1.2$\pm$ 0.2 &  1.4$\pm$ 0.1 &  0.32$\pm$ 0.03 &   1.16   &   1.01   &   0.15   \\ 
HD 13147   &   117.37   &  5502  &   3.94 & 1.0$\pm$ 0.2 &  2.9$\pm$ 0.1 &  0.03$\pm$ 0.02 &   1.11   &   1.02   &   0.09   \\ 
HD 13350   &   108.93   &  5515  &   4.22 & 1.2$\pm$ 0.2 &  2.7$\pm$ 0.1 &  0.25$\pm$ 0.01 &   0.76   &   0.64   &   0.12   \\ 
HD 15507   &    58.79   &  5766  &   4.62 & 1.0$\pm$ 0.2 &  2.9$\pm$ 0.1 &  0.09$\pm$ 0.02 &   1.41   &   1.33   &   0.08   \\ 
HD 23398   &    77.46   &  5592  &   4.10 & 1.2$\pm$ 0.2 &  2.9$\pm$ 0.1 &  0.38$\pm$ 0.02 &   1.48   &   1.37   &   0.11   \\ 
HD 26071   &    90.01   &  5549  &   4.16 & 1.2$\pm$ 0.2 &  2.8$\pm$ 0.1 &  0.12$\pm$ 0.02 &   1.05   &   0.96   &   0.09   \\ 
HD 29231   &    28.10   &  5400  &   4.43 & 1.2$\pm$ 0.2 &  1.9$\pm$ 0.1 &  0.02$\pm$ 0.02 &   0.59   &   0.50   &   0.09   \\ 
HD 38459   &    35.29   &  5233  &   4.43 & 1.2$\pm$ 0.2 &  4.0$\pm$ 0.1 &  0.06$\pm$ 0.02 &   0.87   &   0.70   &   0.17   \\ 
HD 40293   &    73.31   &  5549  &   4.51 & 1.0$\pm$ 0.2 &  1.8$\pm$ 0.1 &  0.00$\pm$ 0.02 &   1.18   &   1.09   &   0.09   \\ 
HD 42719   &    70.57   &  5809  &   4.08 & 1.4$\pm$ 0.2 &  4.4$\pm$ 0.1 &  0.11$\pm$ 0.01 &   1.16   &   1.10   &   0.06   \\ 
HD 42936   &    46.17   &  5126  &   4.44 & 0.8$\pm$ 0.2 &  1.4$\pm$ 0.1 &  0.19$\pm$ 0.02 &   0.96   &   0.78   &   0.18   \\ 
HD 45133   &    62.03   &  5601  &   4.31 & 1.2$\pm$ 0.2 &  2.9$\pm$ 0.1 &  0.16$\pm$ 0.02 &   0.84   &   0.72   &   0.12   \\ 
HD 49866   &    93.46   &  5712  &   3.71 & 1.4$\pm$ 0.2 &  4.7$\pm$ 0.1 & -0.12$\pm$ 0.02 &   1.09   &   1.02   &   0.07   \\ 
HD 50652   &    72.52   &  5641  &   4.21 & 1.2$\pm$ 0.2 &  2.9$\pm$ 0.1 &  0.12$\pm$ 0.02 &   1.15   &   1.07   &   0.08   \\ 
HD 56259   &   112.99   &  5489  &   3.94 & 1.2$\pm$ 0.2 &  3.0$\pm$ 0.1 &  0.11$\pm$ 0.01 &   1.46   &   1.46   &   0.00   \\ 
HD 56413   &    60.75   &  5648  &   4.41 & 1.2$\pm$ 0.2 &  2.9$\pm$ 0.1 &  0.11$\pm$ 0.02 &   1.03   &   0.95   &   0.08   \\ 
HD 61475   &    43.14   &  5250  &   4.47 & 1.2$\pm$ 0.2 &  2.0$\pm$ 0.1 &  0.10$\pm$ 0.02 &   0.81   &   0.65   &   0.16   \\ 
HD 69721   &    47.62   &  5296  &   4.47 & 1.0$\pm$ 0.2 &  1.7$\pm$ 0.1 &  0.14$\pm$ 0.03 &   0.96   &   0.81   &   0.15   \\ 
HD 76849   &    46.82   &  5223  &   5.00 & 1.0$\pm$ 0.2 &  1.9$\pm$ 0.2 & -0.26$\pm$ 0.04 &   0.60   &   0.50   &   0.10   \\ 
HD 78286   &    66.36   &  5794  &   4.40 & 1.2$\pm$ 0.2 &  2.9$\pm$ 0.1 &  0.09$\pm$ 0.02 &   1.25   &   1.18   &   0.07   \\ 
HD 91682   &    82.99   &  5614  &   4.13 & 1.2$\pm$ 0.2 &  3.0$\pm$ 0.1 &  0.08$\pm$ 0.01 &   1.15   &   1.07   &   0.08   \\ 
HD 93849   &    70.27   &  6153  &   4.21 & 1.2$\pm$ 0.2 &  4.3$\pm$ 0.1 &  0.08$\pm$ 0.01 &   0.86   &   0.80   &   0.06   \\ 
HD 95136   &    73.21   &  5744  &   4.41 & 1.2$\pm$ 0.2 &  3.1$\pm$ 0.1 &  0.04$\pm$ 0.01 &   1.20   &   1.13   &   0.07   \\ 
HD 96494   &    49.43   &  5356  &   4.53 & 1.0$\pm$ 0.2 &  2.8$\pm$ 0.1 &  0.06$\pm$ 0.02 &   0.90   &   0.75   &   0.15   \\ 
HD 102196  &    96.34   &  6012  &   3.90 & 1.4$\pm$ 0.2 &  5.9$\pm$ 0.1 & -0.05$\pm$ 0.02 &   1.14   &   1.10   &   0.04   \\ 
HD 106937  &    78.49   &  5455  &   4.05 & 1.2$\pm$ 0.2 &  2.5$\pm$ 0.1 &  0.13$\pm$ 0.02 &   1.05   &   0.94   &   0.11   \\ 
HD 108953  &    67.20   &  5514  &   4.43 & 1.0$\pm$ 0.2 &  2.5$\pm$ 0.1 &  0.25$\pm$ 0.02 &   1.09   &   0.99   &   0.10   \\ 
HD 126535  &    41.67   &  5284  &   4.65 & 1.0$\pm$ 0.2 &  2.1$\pm$ 0.1 &  0.10$\pm$ 0.02 &   0.91   &   0.75   &   0.16   \\ 
HD 149782  &    59.49   &  5554  &   4.34 & 1.2$\pm$ 0.2 &  2.3$\pm$ 0.1 & -0.05$\pm$ 0.02 &   1.03   &   0.94   &   0.09   \\ 
HD 150936  &    79.30   &  5542  &   4.12 & 1.0$\pm$ 0.2 &  4.1$\pm$ 0.2 & -0.03$\pm$ 0.02 &   1.32   &   1.21   &   0.11   \\ 
HD 165204  &    76.16   &  5557  &   4.33 & 1.0$\pm$ 0.2 &  2.7$\pm$ 0.1 &  0.17$\pm$ 0.02 &   1.07   &   0.98   &   0.09   \\ 
HD 170706  &   118.06   &  5698  &   4.40 & 1.0$\pm$ 0.2 &  3.0$\pm$ 0.1 &  0.13$\pm$ 0.02 &   1.26   &   1.18   &   0.08   \\ 
HD 178340  &    46.75   &  5538  &   4.38 & 1.2$\pm$ 0.2 &  2.0$\pm$ 0.1 &  0.16$\pm$ 0.01 &   0.91   &   0.79   &   0.12   \\ 
HD 178787  &    46.77   &  5216  &   4.44 & 1.0$\pm$ 0.2 &  1.6$\pm$ 0.1 &  0.14$\pm$ 0.02 &   0.94   &   0.77   &   0.17   \\ 
HD 185679  &    65.45   &  5681  &   4.43 & 1.0$\pm$ 0.2 &  2.8$\pm$ 0.1 &  0.01$\pm$ 0.02 &   1.29   &   1.20   &   0.09   \\ 
HD 186194  &    86.96   &  5668  &   4.30 & 1.2$\pm$ 0.2 &  2.9$\pm$ 0.1 &  0.07$\pm$ 0.01 &   0.86   &   0.75   &   0.11   \\ 
HD 186265  &    87.26   &  5562  &   4.39 & 1.2$\pm$ 0.2 &  2.5$\pm$ 0.1 &  0.27$\pm$ 0.02 &   1.06   &   0.96   &   0.10   \\ 
HD 193690  &    62.46   &  5558  &   4.48 & 1.0$\pm$ 0.2 &  2.3$\pm$ 0.1 &  0.15$\pm$ 0.02 &   1.13   &   1.04   &   0.09   \\ 
HD 200869  &    62.70   &  5401  &   4.37 & 1.0$\pm$ 0.2 &  1.7$\pm$ 0.1 &  0.25$\pm$ 0.02 &   0.93   &   0.79   &   0.14   \\ 
HD 201757  &    73.91   &  5597  &   4.23 & 1.2$\pm$ 0.2 &  3.7$\pm$ 0.1 &  0.05$\pm$ 0.02 &   1.11   &   1.03   &   0.08   \\ 
HD 218960  &    93.28   &  5732  &   4.27 & 1.2$\pm$ 0.2 &  3.5$\pm$ 0.1 &  0.05$\pm$ 0.01 &   1.21   &   1.14   &   0.07   \\ 
HD 219011  &    85.47   &  5642  &   4.21 & 1.2$\pm$ 0.2 &  3.0$\pm$ 0.1 &  0.13$\pm$ 0.02 &   1.15   &   1.07   &   0.08   \\ 
HD 219556  &    59.21   &  5485  &   4.44 & 1.0$\pm$ 0.2 &  2.0$\pm$ 0.1 &  0.04$\pm$ 0.02 &   1.03   &   0.93   &   0.10   \\ 
HD 220981  &    62.58   &  5567  &   4.33 & 1.0$\pm$ 0.2 &  2.4$\pm$ 0.1 &  0.11$\pm$ 0.02 &   1.07   &   0.98   &   0.09   \\ 
HD 221575  &    32.79   &  5037  &   4.49 & 1.4$\pm$ 0.2 &  3.3$\pm$ 0.1 & -0.11$\pm$ 0.03 &   0.98   &   0.80   &   0.18   \\ 
HIP 28641  &    90.33   &  5747  &   4.45 & 1.0$\pm$ 0.2 &  3.0$\pm$ 0.1 & -0.01$\pm$ 0.02 &   1.41   &   1.33   &   0.08   \\ 
HIP 29442  &    76.22   &  5322  &   4.43 & 0.8$\pm$ 0.2 &  1.8$\pm$ 0.1 &  0.26$\pm$ 0.02 &   0.91   &   0.75   &   0.16   \\ 
HIP 43267  &    79.55   &  5642  &   4.29 & 1.2$\pm$ 0.2 &  2.6$\pm$ 0.1 &  0.12$\pm$ 0.02 &   1.41   &   1.31   &   0.10   \\ 
HIP 51987  &    85.25   &  6158  &   5.10 & 1.0$\pm$ 0.2 &  2.8$\pm$ 0.1 &  0.27$\pm$ 0.02 &   1.52   &   1.48   &   0.04   \\ 
HIP 53084  &    60.98   &  5527  &   4.32 & 1.0$\pm$ 0.2 &  2.6$\pm$ 0.1 &  0.25$\pm$ 0.02 &   1.11   &   1.01   &   0.10   \\ 
HIP 57331  &    79.49   &  5531  &   4.21 & 1.2$\pm$ 0.2 &  2.8$\pm$ 0.1 &  0.09$\pm$ 0.01 &   1.47   &   1.37   &   0.10   \\ 
HIP 66990  &    72.99   &  5595  &   4.15 & 1.2$\pm$ 0.2 &  2.8$\pm$ 0.1 &  0.16$\pm$ 0.02 &   1.08   &   1.00   &   0.08   \\ 
HIP 69724  &    67.29   &  5793  &   4.79 & 1.0$\pm$ 0.2 &  2.5$\pm$ 0.1 &  0.28$\pm$ 0.02 &   1.54   &   1.47   &   0.07   \\ 
HIP 111286 &    97.37   &  5690  &   4.19 & 1.2$\pm$ 0.2 &  3.0$\pm$ 0.1 &  0.07$\pm$ 0.02 &   1.49   &   1.36   &   0.13   \\ 
\end{tabular}
\end{table*}

\clearpage

\begin{longtable}[c]{cccc}                    
\caption{Full list of stars and their NLTE and LTE Li abundances. Upper limits are maked by itallic fonts. \label{_age}} \\
\hline\hline
       &      Name      &  \NLTE   &   \LTE   \\\hline           
   1   &   HD 6790      &   2.51   &   2.48   \\                 
   2   &   HD 7950      &   1.65   &   1.54   \\                 
   3   &   HD 8389      & {\it  1.16}   & {\it  1.01}   \\       
   4   &   HD 8446      &   2.70   &   2.66   \\                 
   5   &   HD 9174      &   1.44   &   1.33   \\                 
   6   &   HD 10188     &   2.79   &   2.75   \\                 
   7   &   HD 10278     &   1.67   &   1.60   \\                 
   8   &   HD 13147     &  {\it 1.11}   & {\it 1.02}   \\        
   9   &   HD 13350     &  {\it 0.76}   & {\it 0.64}   \\        
  10   &   HD 15507     &  {\it 1.41}   & {\it 1.33}   \\        
  11   &   HD 18708     &   1.89   &   1.82   \\                 
  12   &   HD 18754     &   2.35   &   2.26   \\                 
  13   &   HD 19493     &   2.03   &   1.96   \\                 
  14   &   HD 19773     &   2.91   &   2.90   \\                 
  15   &   HD 23398     &  {\it 1.48}   &  {\it 1.37}   \\       
  16   &   HD 26071     &  {\it 1.05}   &  {\it 0.96}   \\       
  17   &   HD 29231     &  {\it 0.59}   &  {\it 0.50}   \\       
  18   &   HD 38459     &  {\it 0.87}   &  {\it 0.70}   \\       
  19   &   HD 38467     &   1.97   &   1.89   \\                 
  20   &   HD 40293     &  {\it 1.18}   &  {\it 1.09}   \\       
  21   &   HD 42538     &   2.61   &   2.59   \\                 
  22   &   HD 42719     &  {\it 1.16}   &  {\it 1.10}   \\       
  23   &   HD 42936     &  {\it 0.96}   &  {\it 0.78}   \\       
  24   &   HD 45133     &  {\it 0.84}   &  {\it 0.72}   \\       
  25   &   HD 48265     &   2.47   &   2.40   \\                 
  26   &   HD 49866     &  {\it 1.09}   &  {\it 1.02}   \\       
  27   &   HD 50652     &  {\it 1.15}   &  {\it 1.07}   \\       
  28   &   HD 55524     &   2.21   &   2.13   \\                 
  29   &   HD 56259     &  {\it 1.46}   &  {\it 1.35}   \\       
  30   &   HD 56413     &  {\it 1.03}   &  {\it 0.95}   \\       
  31   &   HD 56957     &   2.25   &   2.17   \\                 
  32   &   HD 61475     &  {\it 0.81}   &  {\it 0.65}   \\       
  33   &   HD 66653     &   1.95   &   1.88   \\                 
  34   &   HD 69721     &  {\it 0.96}   &  {\it 0.81}   \\       
  35   &   HD 76849     &  {\it 0.60}   &  {\it 0.50}   \\       
  36   &   HD 77338     &  {\it 0.80}   &  {\it 0.65}   \\       
  37   &   HD 78130     &   1.97   &   1.90   \\                 
  38   &   HD 78286     &  {\it 1.25}   &  {\it 1.18}   \\       
  39   &   HD 86006     &   1.92   &   1.83   \\                 
  40   &   HD 90028     &   2.54   &   2.48   \\                 
  41   &   HD 90520     &   2.55   &   2.55   \\                 
  42   &   HD 91682     &  {\it 1.15}   &  {\it 1.07}   \\       
  43   &   HD 93849     &  {\it 0.86}   &  {\it 0.80}   \\       
  44   &   HD 95136     &  {\it 1.20}   &  {\it 1.13}   \\       
  45   &   HD 96494     &  {\it 0.90}   &  {\it 0.75}   \\       
  46   &   HD 101197    &   2.20   &   2.13   \\                 
  47   &   HD 101348    &   2.50   &   2.43   \\                 
  48   &   HD 102196    &  {\it 1.14}   &  {\it 1.10}   \\       
  49   &   HD 102361    &   2.72   &   2.74   \\                 
  50   &   HD 105750    &   2.37   &   2.30   \\                 
  51   &   HD 106937    &  {\it 1.05}   &  {\it 0.94}   \\       
  52   &   HD 107181    &   1.91   &   1.81   \\                 
  53   &   HD 108953    &  {\it 1.09}   &  {\it 0.99}   \\       
  54   &   HD 126535    &  {\it 0.91}   &  {\it 0.75}   \\       
  55   &   HD 127423    &   2.72   &   2.70   \\                 
  56   &   HD 128356    &  {\it 0.61}   &  {\it 0.43}   \\       
  57   &   HD 143120    &   2.34   &   2.25   \\                 
  58   &   HD 143361    &  {\it 0.95}   &  {\it 0.83}   \\       
  59   &   HD 144550    &   2.35   &   2.27   \\                 
  60   &   HD 144848    &   1.86   &   1.78   \\                 
  61   &   HD 144899    &   2.69   &   2.65   \\                 
  62   &   HD 147873    &  {\it 0.63}   &  {\it 0.60}   \\       
  63   &   HD 149189    &   2.34   &   2.28   \\                 
  64   &   HD 149782    &  {\it 1.03}   &  {\it 0.94}   \\       
  65   &   HD 150936    &  {\it 1.32}   &  {\it 1.21}   \\       
  66   &   HD 152079    &  {\it 1.29}   &  {\it 1.20}   \\       
  67   &   HD 154221    &   1.78   &   1.71   \\                 
  68   &   HD 154672    &  {\it 1.12}   &  {\it 1.04}   \\       
  69   &   HD 158469    &   2.58   &   2.55   \\                 
  70   &   HD 165155    &  {\it 1.07}   &  {\it 0.96}   \\       
  71   &   HD 165204    &  {\it 1.07}   &  {\it 0.98}   \\       
  72   &   HD 170706    &  {\it 1.26}   &  {\it 1.18}   \\       
  73   &   HD 178340    &  {\it 0.91}   &  {\it 0.79}   \\       
  74   &   HD 178787    &  {\it 0.94}   &  {\it 0.77}   \\       
  75   &   HD 185679    &  {\it 1.29}   &  {\it 1.20}   \\       
  76   &   HD 186194    &  {\it 0.86}   &  {\it 0.75}   \\       
  77   &   HD 186265    &  {\it 1.06}   &  {\it 0.96}   \\       
  78   &   HD 189627    &   3.01   &   3.01   \\                 
  79   &   HD 190125    &   1.71   &   1.63   \\                 
  80   &   HD 191122    &   1.90   &   1.83   \\                 
  81   &   HD 191760    &   2.42   &   2.36   \\                 
  82   &   HD 193690    &  {\it 1.13}   &  {\it 1.04 }  \\       
  83   &   HD 193995    &   1.58   &   1.50   \\                 
  84   &   HD 194490    &   1.76   &   1.70   \\                 
  85   &   HD 200869    &  {\it 0.93}   &  {\it 0.79}   \\       
  86   &   HD 201757    &  {\it 1.11}   &  {\it 1.03}   \\       
  87   &   HD 206683    &   2.25   &   2.19   \\                 
  88   &   HD 206837    &   2.26   &   2.18   \\                 
  89   &   HD 218960    &  {\it 1.21}   &  {\it 1.14}   \\       
  90   &   HD 219011    &  {\it 1.15}   &  {\it 1.07}   \\       
  91   &   HD 219556    &  {\it 1.03}   &  {\it 0.93}   \\       
  92   &   HD 220981    &  {\it 1.07}   &  {\it 0.98}   \\       
  93   &   HD 221575    &  {\it 0.98}   &  {\it 0.80}   \\       
  94   &   HD 221954    &   1.57   &   1.48   \\                 
  95   &   HD 222910    &   1.54   &   1.43   \\                 
  96   &   HD 224538    &   2.75   &   2.73   \\                 
  97   &   HIP 19807    &   1.82   &   1.76   \\                 
  98   &   HIP 28641    &  {\it 1.41}   &  {\it 1.33}   \\       
  99   &   HIP 29442    &  {\it 0.91}   &  {\it 0.75}   \\       
 100   &   HIP 31831    &   2.41   &   2.35   \\                 
 101   &   HIP 43267    &  {\it 1.41}   &  {\it 1.31}  \\        
 102   &   HIP 51987    &  {\it 1.52}   &  {\it 1.48}   \\       
 103   &   HIP 53084    &  {\it 1.11}   &  {\it 1.01}   \\       
 104   &   HIP 57331    &  {\it 1.47}   &  {\it 1.37}   \\       
 105   &   HIP 66990    &  {\it 1.08}   &  {\it 1.00}   \\       
 106   &   HIP 69724    &  {\it 1.54}   &  {\it 1.47}   \\       
 107   &   HIP 111286   &  {\it 1.49}   &  {\it 1.36}   \\       
\hline
\end{longtable}

\end{appendix}
\end{onecolumn}

\end{document}